\documentclass[manuscript]{aastex}

\usepackage{natbib}
\usepackage{graphicx}
\usepackage{longtable}
\usepackage{array}   
\usepackage{lscape}
\usepackage{multirow}
\usepackage[ampersand]{easylist}
\usepackage{color}

\newcommand{\RNum}[1]{\uppercase\expandafter{\romannumeral #1\relax}}
\newcolumntype{D}{ >{\centering\arraybackslash} m{4.3cm} }
\newcolumntype{G}{ >{\centering\arraybackslash} m{1.0cm} }
\newcolumntype{H}{ >{\centering\arraybackslash} m{2.0cm} }

\newcolumntype{S}{ >{\centering \arraybackslash} m{1.5cm} }
\newcolumntype{T}{ >{\centering \arraybackslash} m{6.1cm} }
\newcolumntype{W}{ >{\centering \arraybackslash} m{1.8cm} }

\newcommand{\changes}[1]{#1}                         

\shorttitle{Analysis of Flows Inside Prominences}
\shortauthors{Freed et al.}

\begin{document}

\title{Analysis of Flows Inside Quiescent Prominences as Captured by Hinode/Solar Optical Telescope}

%
\author{M.S. Freed, D.E. McKenzie and D.W. Longcope}
\affil{Physics Department, Montana Sate University,
    Bozeman, MT 59717}
\email{mfreed@physics.montana.edu}

\author{M. Wilburn}
\affil{Transylvania University, Lexington, KY 40508}


\begin{abstract}
Developing an understanding of how magnetic fields can become entangled in a prominence is important for predicting a possible eruption. This work investigates the kinetic energy and vorticity associated with plasma motion residing inside quiescent prominences. These plasma flow characteristics can be utilized to improve our understanding of how the prominence maintains a stable magnetic field configuration. Three different contrast-enhanced solar prominence observations from \textit{Hinode}/Solar Optical Telescope were used to construct velocity maps -- in the plane of the sky -- via a Fourier local correlation tracking program. The resulting velocities were then used to perform the first ever analysis of the two-dimensional kinetic energy and enstrophy spectra of a prominence. Enstrophy is introduced here as a means of quantifying the vorticity which has been observed in many quiescent prominences. The kinetic energy power spectral density produced indices ranging from -1.00 to -1.60. There was a consistent anisotropy in the kinetic energy spectrum of all three prominences examined. Examination of the intensity power spectral density reveals that a different scaling relationship exists between the observed prominence structure and velocity maps. All of the prominences exhibited an inertial range of at least $0.8 \leq k\leq 2.0\; \textrm{rads} \: \textrm{Mm}^{-1}$. Quasi-periodic oscillations were also detected in the centroid of the velocity distributions for one prominence. Additionally, a lower limit was placed on the kinetic energy density ($\epsilon \, \sim 0.22-7.04\:  \mathrm{km}^{2}\textrm{s}^{-2}$) and enstrophy density ($\omega \, \sim 1.43-13.69\: \times 10^{-16} \, \textrm{s}^{-2}$) associated with each prominence.

\end{abstract}



%

\section{Introduction}

Quiescent prominences (QPs) form far from active regions and also go by the name ``polar crown" prominence when they appear at higher latitudes. These structures form a canopy of plasma over magnetic polarity inversion lines (PIL) and are surrounded by a region of depleted emission referred to as coronal cavities. Prominences consist of cool mass extending well into the corona -- far beyond a scale height. QPs are also remarkably stable and can have lifetimes that last for over a month. However, some QPs can eventually lose stability for reasons we do not fully understand. In early models \citep{Kippenhahn57,Kuperus74} the mass was supported in static equilibrium. Observations seem to reveal, however, continuous motion of the material. The prominence mass is actually in a dynamic state, which consists of plasma precipitating down toward the photosphere and being disturbed by intrusive voids coming up from below. A prominence is seen as a bright optically thin plasma when viewed on the solar limb with H$\alpha$. However, the same structure appears dark when seen on the solar disk in H$\alpha$. This subtle difference in appearance is denoted by referring to prominences as filaments when viewed against the photosphere. A more comprehensive review of prominences can be found in \cite{Vial15} and \cite{Parenti14}. 

All of the prominences for this study possess a ``hedgerow" topology. They consist of large curtains of vertical threads with arches or cavities forming at the base. These hedgerow prominences have been observed to undergo downflows \citep{Engvold81,Martres81} and cyclone motion \citep{Liggett84,Berger08} inside the vertical curtain from ground base observatories. The Solar Optical Telescope (SOT) \citep{Tsuneta08} aboard the \textit{Hinode} spacecraft \citep{Kosugi07} provides prominence observations that clearly show upflow motions.\par

Previous investigations with \textit{Hinode}/SOT have shown cavities or ``bubbles" forming next to the chromosphere and columnar ``plume" structures rising from these bubbles into the prominence \citep{Berger08,Berger10}. A similar evolution was also witnessed by Mauna Loa Solar Observatory (MLSO), though with less sharp spatial resolution \citep{Detoma08}. The plume typically begins as a slight perturbation around the prominence bubble and gradually ascends. Next, the plume becomes disconnected from the underlying bubble after 2-3 minutes; it continues up toward the high altitude prominence sheet and eventually becomes lost among downward \& vortex flows at its apex. Plumes have been measured to have a maximum width of 2-6 Mm, lifetimes of 5-15 minutes, and ascend to a height of 15 Mm with an average velocity of 13-17 km s$^{-1}$ \citep{Berger10}. Movies of this activity show downward flows surrounding the rising plume, which acts to slow its ascending speed. The plumes appear to shed Kelvin-Helmholtz vortex formations from their boundaries as they ascend \citep{Berger10}. The initial perturbation that generates plumes from the prominence bubble is hypothesized to be a Rayleigh-Taylor instability. A magneto-thermal convection system is created and driven by the prominence bubble, which has plasma that is 25-120 times hotter than the (overlaying) prominence \citep{Berger11}.\par

Modeling by \cite{Low95} has shown that the region between the photosphere and the twisted-horizontal magnetic fields found at the center of the prominence has a very weak field strength. This means the plumes can start moving in the peripheral region of the prominence where the buoyant force is dominating over the Lorentz force, i.e., a high $\beta$. (The $\beta$ parameter is simply the ratio of plasma pressure to magnetic pressure.) Similar claims of high beta (0.391$\leq \beta \leq$ 1.210) are reported by \cite{Hillier12}, who compared numerical simulations of rising plumes in a modified Kippenhahn-Schluter prominence model with observations. The magnetic field would therefore become more twisted and entangled by the turbulent flow, potentially leading to destabilizing internal reconnection \citep{Ball08}. \par

Work by \cite{Haerendel11} illustrates one possible explanation for how vertical pillared structures can exist inside the predominately horizontal magnetic field of prominences. They propose that packets of plasma can collect and are then squeezed through the horizontal field due to gravity \citep[See Figure 2 of][]{Haerendel11}. The downward flow is then counter balanced by a friction force due to dissipation of Alfven waves along the horizontal magnetic field, which can also explain the constant motion seen in observations. Larger vertical features have also been produced when modeling the three-dimensional global evolution of prominences \citep{Terradas15}.

There are still many outstanding questions that need to be addressed in our understanding of prominences and filaments. What causes the short timescale changes to the magnetic field configuration that result in coronal mass ejections? Are the observed dynamics altering the energy stored in the magnetic field in a significant way? How can the plasma in these prominences be maintained in the corona for so long? What does the three-dimensional velocity profile look like inside these prominences? To answer these questions, we need to start examining the temporal evolution of the entire prominence.\par

The objective of this study is to quantify the turbulent velocity and vorticity structure inherit to the motion found in QPs. Velocity measurements for the plasma residing inside prominences or filaments can be determined by Doppler velocity measurements for motion parallel to the line of sight \citep{Engvold78,Lin03,Schmieder14} and by tracking intensity features in images for motion normal to the line of sight \citep{Engvold76,Zirker94,Chae08b,Berger10}. \par

We investigate the motion in the plane of the sky by using local correlation tracking to determine the velocity at each pixel in the intensity images. The structure of this derived motion can be determined either by using either structure functions or by calculating the power spectral density (PSD). One-dimensional versions of PSDs have been measured for intensity images associated with the prominence on 2006/11/30 by \cite{Leonardis12}. The present work extends that investigation by measuring the PSDs associated with velocity, vorticity, and intensity in two dimensions.\par

PSDs give an indication of the amount of power present in each spatial or temporal frequency in Fourier space. The PSD can then be integrated over all frequencies to give the total power of the signal. The resulting spectra can also be used to establish if a power-law relationship exists in the velocity, vorticity, or intensity. Such a relationship would mean these quantities are scale invariant, or self-similar at different length scales. This kind of analysis on motion has been done before in the high corona for solar wind \citep{Podesta06,Boldyrev11,Bourouaine12} and in the photosphere \citep{Abramenko05,Rieutord10,Matsumoto10}. However, we are unaware of analyses of PSDs for motion inside prominences. Therefore, in this article we describe PSDs of kinetic energy (per unit of mass), as a means of characterizing the velocities, including a measure of anisotropy in the same. In order to investigate how the observed vorticity changes in prominence observations \citep{Liggett84,Berger08}, including the distribution of vorticity across a range of length scales, we also calculate PSDs of the enstrophy.\par

Enstrophy is the integral of the squared vorticity at a given location. A driving force causes large eddies, which then cascade down to smaller and smaller eddies. This process offers a means of transferring energy and momentum down to the dissipation scale, which can be used for increasing the magnetic reconnection rate in plasmas with low magnetic diffusivity. We investigate whether the PSDs associated with enstrophy, kinetic energy, and image intensity show evidence for a power-law distribution with length scales. If they do show self-similarity, what limits can be found for the inertial range for this behavior?\par 

In Section 2, we describe the observations taken with \textit{Hinode}/SOT and then how the data were reduced for our analysis. An explanation of how the velocity fields were determined for each prominence can be found in Section 3. The method used for calculating the one-dimensional power spectral density from the two-dimensional data sets is explained in Section 4, and all of the results are described in Section 5. A discussion on the characteristics measured can be found in Section 6, with concluding remarks appearing in Section 7.\par

\section{Observations and Data Reduction}

Three different data sets are used for this study, and all of them were obtained from the JAXA/NASA jointly operated \textit{Hinode} spacecraft. \textit{Hinode} contains three major systems onboard: the Solar Optical Telescope (SOT), X-Ray Telescope (XRT), and EUV Imaging Spectrometer (EIS). SOT is based upon an Optical Telescope Assembly (OTA) that houses a Gregorian telescope with 50 cm primary mirror aperture. Its Focal Plane Package (FPP) has three instruments, the Broadband Filter Imager (BFI), Narrowband Filter Imager (NFI), and Spectro-Polarimeter (SP) \citep{Kosugi07,Tsuneta08}. The research presented here uses data taken from the BFI and NFI. Information pertaining to the cadence, filter used, spatial resolution, and time span associated with each prominence observation can be found in Table \ref{table:A}, along with its Solar Object Locator (SOLyyyy-mm-ddThh:mm:ssLdddCddd, where T is the universal time, L is the Carrington longitude, and C is the co-latitude). The Ca \RNum{2} H-line ($\lambda=396.85\;nm$) data come from the BFI, which has a ${218}''\times{109}''$ field of view. The H$\alpha$ ($\lambda=656.3\;nm$) data come from SOT's tunable Lyot filter NFI with a field of view of ${328}''\times{164}''$. After binning the pixels $2\times2$ to conserve spacecraft telemetry, the Ca \RNum{2} images have an angular resolution of ${0.109}''$ per CCD pixel and the H$\alpha$ images have a pixel scale of ${0.16}''$. The SOT data were prepared with the SolarSoft IDL \citep{freeland98} \texttt{fg\_prep} program, which removed dark current, analog-to-digital converter offsets, cosmic ray spikes, and corrected for known camera readout defects in the images. \par

SOT uses a closed-loop Correlation Tracker (CT) system with a field of view of $11''\times11''$ to track displacements in solar granulation motion, and then makes corrections via a piezo-driven tip-tilt mirror to reduce the amount of spacecraft jitter seen in solar images \citep{Shimizu08}. However, the CT has difficulty maintaining its tracking at the solar limb and therefore images will experience a sub-pixel drift over a period of several minutes \citep{Berger10}. This drift was found to vary with a maximum displacement between 4 and 8 pixels over a time period 70-135 minutes for the Ca \RNum{2} data sets on 2006/11/30 and 2007/10/03. The H$\alpha$ data on 2007/04/25 reached a maximum drift of 15-20 pixels over the 147.5 minutes data set. It should be noted that the image-to-image drift did not vary by more than a few tenths of a pixel. However, these effects were reduced further by applying the cross-correlation SolarSoft IDL program \texttt{tr\_get\_disp} to the data in 5-minute increments to find and correct for this slow drift. Small groups consisting of only 5 minutes worth of data were used, because larger time spans caused the cross-correlating program to produce results that were too jittery for further analysis. Any velocities derived between any two consecutive 5 minute groups with the local correlation tracking program, described in Section \ref{velocity_measurements_section}, were completely removed from the data cube to avoid any possible residual jitter caused by improper co-alignment.\par

Next, all of the data were rotated to place the solar limb parallel to the bottom of the image. The images were then contrast enhanced in order for the fainter plasma on the limb to be seen as well as the more intense solar disk material. Several different contrast enhancement techniques were explored that included histogram equalization and applying a function to image intensity values in order to stretch out the intensity range. The latter method can be done by multiplying the image array by some exponent or by applying a more involved function like a Gaussian for observing the fainter features in the solar corona.\par

\begin{figure}
\centering

   \includegraphics[scale=0.70]{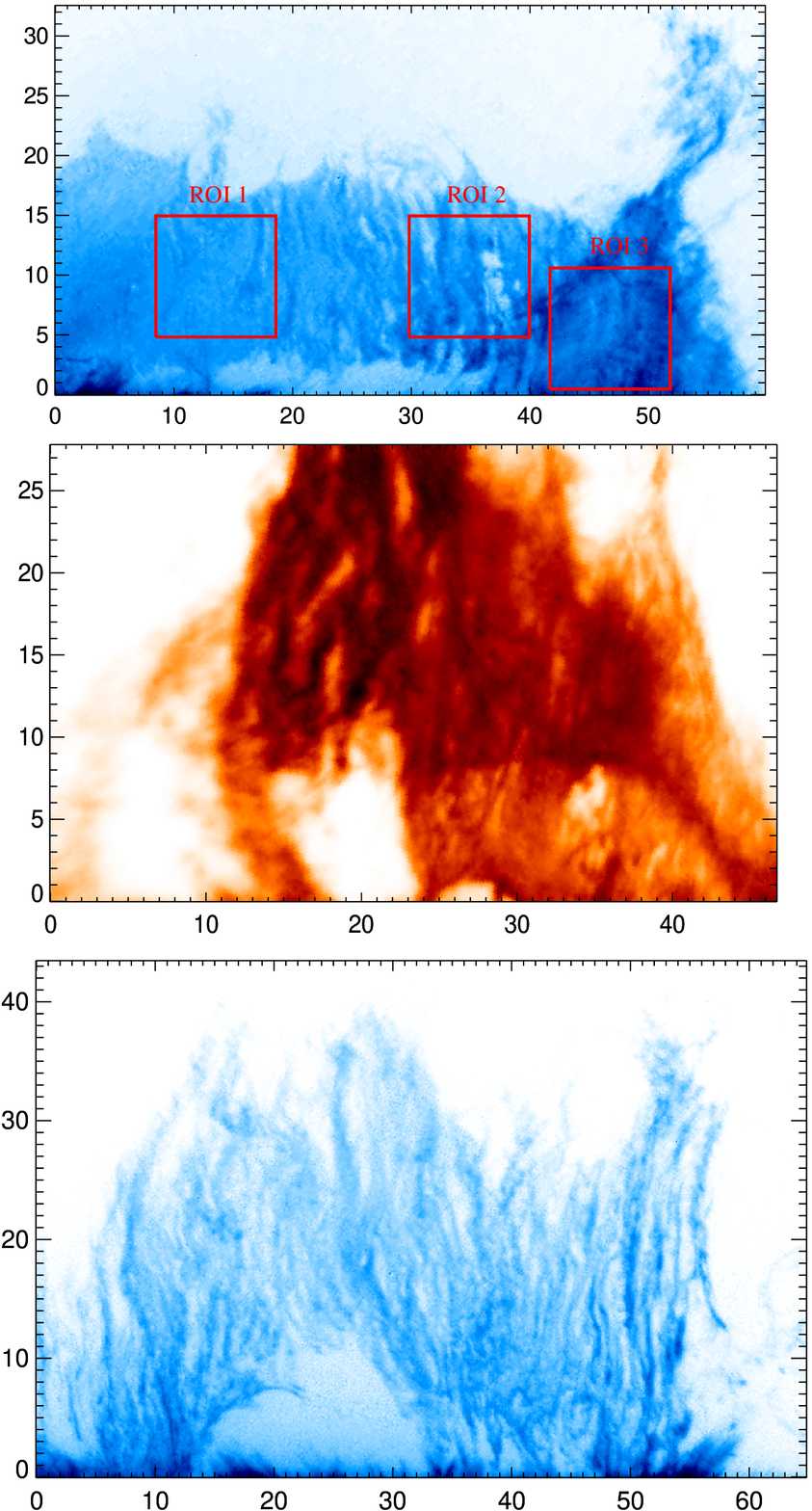}\\

  \caption{Images of the contrast enhanced and unsharp masked solar prominence on 2006/11/30 06:29:22 UT (top), 2007/04/25 13:29:06 UT (middle), and 2007/10/03 02:46:16 UT (bottom). All of the spatial units are in Mm. See the electronic version online for the corresponding intensity movies.}
  \label{intensity_figure}
\end{figure}

 We settled on the ``asinh method'' developed by \cite{Lupton99}, which uses the inverse hyperbolic sine function to improve contrast. (Incidentally, Lupton's original intent for this method was to improve the detection of faint objects with low signal-to-noise ratios in the Sloan Digital Sky Survey. Now it is being implemented here for detecting the dimmest areas surrounding the brightest object in the sky.) The inverse hyperbolic sine function can be written as $\mathrm{sinh}^{-1}\left ( x/\beta \right )$, where $\beta$ is a ``softening parameter" that has been applied to the intensity at each pixel. This function has the advantage of contrast enhancing faint image areas ($\left | x \right |\ll \beta$) linearly and bright regions ($\left | x \right |\gg  \beta$) logarithmically. $\beta$ is used for specifying the pivot point where the scaling behavior switches, and was determined by visually inspecting the images for the optimum value that reduced the unwanted low signal-to-noise regions surrounding the prominence. The minimum and maximum intensity values were also cropped to improve the final image contrast range. The removal of the low signal-to-noise region off the limb was crucial for eliminating residual noise that could introduce unreliable velocities to plasma flows as described in Section \ref{velocity_measurements_section}. A complete list of parameters used for contrast enhancing each prominence can be found in Table \ref{table:B}. Figure \ref{intensity_figure} illustrates the effect of the contrast enhancement on the 2006/11/30 06:26:48 UT (top), 2007/04/25 13:27:07 UT (middle), and 2007/10/03 02:46:16 UT (bottom) prominences, with an additional unsharp mask applied to make features easier to see in figures. The reader can also see intensity movies by visiting the electronic version of this paper online.\par

Memory constraints required us to split our full data cubes into smaller cubes.  It should be noted that there was a gap in the SOT data for the 2006/11/30 prominence of 15 minutes at 05:45 UT, and also a gap of 5 minutes at 03:30 UT for the 2007/10/03 prominence. The data gaps provide natural break points at which to subdivide the data for analysis. Then, in the cases of 2006/11/30 and 2007/04/25, we further broke the shortened intervals into sets of even and odd frames, so that smaller minimum velocities could be detected by the local correlation tracking algorithm. It was determined, from inspection of advected corks described in Section \ref{velocity_measurements_section}, that at least 30 seconds was needed between SOT images in order for FLCT to detect variations in the intensities. The supplemental intensity movies show the results with all the images for the 2006/11/30 and 2007/04/25 prominences included. \par

\section{Velocity Measurements}
\label{velocity_measurements_section}

Velocities were obtained from the contrast enhanced images with the use of Fisher \& Welsch's Fourier Local Correlation Tracking program \citep[FLCT:][]{Fisher08}. The FLCT method does however have some limitations that are inherent to any local correlation tracking program. It assumes that each pixel is surrounded by other pixels with nearly the same velocity. Any motions that are normal to the intensity gradient are difficult to detect. There are also problems with trying to detect velocities at small spatial scales and of course with determining velocities when multiple features are present along the line of sight. However, this technique does have the advantage of running quicker than other optical flow codes available, as can be seen in Figure 2 of \cite{Chae08}.  \changes{Any variation in the intensity at a pixel is assumed to be associated with plasma motion and not the result of temperature changes in the prominence. It is conceivable that some of this variation can be attributed to changes in emission measure as well. Temperature changes can be a factor by moving plasma in and out of the filter bandpass. The contributions of both factors will be explored further in future work, but are expected to be small, based on preliminary analysis.} All results are then verified, as outlined by \cite{Mckenzie13}, by placing test particles (corks) into the derived velocity field, with the original intensity image in the background, and advecting the corks by the FLCT velocities. This offers a means for determining if the derived motions resemble the motions appearing in the intensity movies.\par

FLCT requires three parameters: $\sigma$ for specifying the size of the apodization window used for calculating the cross-correlation; minimum intensity threshold ($t$) that has been normalized to the maximum absolute value of the image averages; and $k$ which is used for applying a low-pass filter in wavenumber space to reduce high frequency noise. The parameter values used for the three prominences are given in Table \ref{table:B}. The FLCT algorithm delivers a displacement vector at each pixel, from which we may determine a velocity.  In some cases no displacement is found due to the intensity threshold parameter ($t$); these locations with no data were assigned a velocity value of zero. The FLCT program also keeps track of these locations by producing a Boolean map, which indicates if the pixel intensity met the minimum cutoff set by $t$. Next, any velocities with an absolute value greater than a conservative acoustic cutoff estimate of 40 km s$^{-1}$ were assumed to be spurious and replaced with the median of the eight spatially adjacent pixels. This procedure is done for all the qualifying pixels once and then repeated N times, where N is the number of adjacent pixels that also have an absolute value greater than the acoustic limit. The number of pixels with velocities above the threshold was small enough -- 0.03\% or less per data set -- as to not affect the resulting power-law indices calculated. \changes{Presently, there is no ab initio estimate of the error associated with the velocities obtained from the FLCT code and therefore can not be propagated through our results. However, Section \ref{Appendix} attempts to quantify how some factors might contribute to the uncertainty associated with the FLCT velocities. For the prominence observed on 2006/11/30, we estimate the error in the FLCT velocity components to be approximately 2.45$\pm$0.30 km s$^{-1}$. But this value overestimates the error at low velocity magnitudes and underestimates it at high velocities, as explained later in Section \ref{Appendix}. }\par

The Boolean maps were then used with a Laplacian filter to determine the location of edges in the derived FLCT velocity field. The velocity fields were then smoothed with a 3x3 boxcar at all edges to prevent any edge effects when taking the Fourier transform of the velocities in our analysis later.\par

A histogram showing the horizontal and vertical velocity distribution for the 2006/11/30 prominence at 06:07:50 UT is shown in Figure \ref{histogram_of_velocities} with bin sizes of 0.33 km s$^{-1}$. (The term ``horizontal" is used in this study for referring to motion parallel to solar surface, while ``vertical" indicates motion normal to the photosphere.) These histograms show a central Gaussian of standard deviation $\sim 2.5$ km s$^{-1}$, surrounded by non-Gaussian wings extending to about 30 km s$^{-1}$. An example of the velocities obtained from the FLCT is illustrated in Figure \ref{figure_of_velocity_map}, which corresponds with the intensity image of the 2006/11/30 prominence in Figure \ref{intensity_figure}.\par

Additionally, the temporal-averages of the velocity fields were calculated from the root mean square velocity ($V_{rms}$) of each time step in the horizontal and vertical direction. The $V_{rms}$ values were calculated by excluding all the missing data points, from each data cube. The standard deviations and temporal-averages are listed in Table \ref{table:A}. \par

\section{Calculating the One Dimensional PSDs}

The power spectral density (PSD) is used to quantify the distribution of image intensities, kinetic energy, or enstrophy, across a range of spatial scales. For simplicity the two-dimensional spectra are angle averaged to yield a function of wavenumber magnitude, $k$. We begin by considering the discrete Fourier spectrum of a single velocity component $v(x,y)=v(\mathbf{x})$, within a rectangular region $(x,y) \in (L_{x}, L_{y})$, sampled on a uniform grid

\begin{equation}
\left \langle \left | \tilde{v}_{\mathbf{k}} \right |^{2} \right \rangle=\frac{1}{N_{x}N_{y}}\sum_{\Delta \mathbf{x}}\left \langle v(\mathbf{x})v(\mathbf{x} + \Delta \mathbf{x}) \right \rangle e^{-i\mathbf{k}\cdot \Delta \mathbf{x}}.
\label{fourier_spectrum}
\end{equation}
A grid of $N_{x}\times N_{y}$ points in position space corresponds to $\mathbf{x}=(x_{m},y_{n})=(m L_{x}/N_{x},n L_{y}/N_{y})$
with a range of $\left [ -(L_{x}/2)\leq x_{m}\leq (L_{x}/2), -(L_{y}/2)\leq y_{n}\leq (L_{y}/2)  \right ]$. Due to the assumed periodicity of region $\mathbf{x}$ in position space, the wavenumbers fall on a conjugate grid $\mathbf{k}=(k_{x},k_{y})=(2\pi m/L_{x},2\pi n/L_{y})$ with the range $\left [ -(\pi N_{x}/L_{x}) < k_{x}< +(\pi N_{x}/L_{x}) ,  -(\pi N_{y}/L_{y}) < k_{y}< +(\pi N_{y}/L_{y})  \right ]$. The quantity in the angled brackets denotes the correlation function of the velocities in position space.\par

Equation (\ref{fourier_spectrum}) can be used to express the power per unit wavenumber (i.e., PSD) in two dimensions as
\begin{equation}
P(\mathbf{k})=\frac{1}{\Delta k_{x}\Delta k_{y}}\left \langle \left | \tilde{v}_{\mathbf{k}} \right |^{2} \right \rangle=\frac{L_{x}L_{y}}{(2\pi)^{2}}\left \langle \left | \tilde{v}_{\mathbf{k}} \right |^{2} \right \rangle.
\label{P:defined}
\end{equation}
Summing $P(\mathbf{k})$ over the entire $\mathbf{k}$ lattice gives
\begin{equation}
\int P(\mathbf{k})dk_{x}dk_{y}\simeq \sum_{\mathbf{k}}\Delta k_{x}\Delta k_{y}P(\mathbf{k})=\sum_{\mathbf{k}}\left \langle \left | \tilde{v}_{\mathbf{k}} \right |^2 \right \rangle=\left \langle v^{2}(\mathbf{x}) \right  \rangle,
\label{eq:C}
\end{equation}
which is the mean energy according to Parseval's theorem. If the statistics associated with $v(x,y)$ were approximately isotropic, then the PSD would depend only on wave-vector magntiude, $P(\mathbf{k})=P(k)$. Even when the velocity is not homogeneous, we can average the function $P({\bf k})$ over angle to arrive at a one-dimension function $P(k)$.  This presents information about scales in a relatively simple form, at the expense of information about anisotropy.  We return in Section \ref{Discussion_section} to analyze the degree of anisotropy in our data.\par

Following the foregoing logic we can either integrate or average over the wave-vector angle to arrive at our isotropized PSD $P(k)$. This is done by grouping wavenumber grid points into annuli $A_{j}$ of mean radius $k_{j}$ and width of $\Delta K_{j}$
\begin{equation}
A_{j}=\left \{ k_{j}-\frac{1}{2}\Delta K_{j} < \left | \mathbf{k} \right | < k_{j}+\frac{1}{2}\Delta K_{j} \right \}.
\end{equation}
Each annulus $A_{j}$ has an area approximately equal to $2\pi k_{j} \Delta K_{j}$ and contains $N_{j}$ wavevectors.
Equation (\ref{eq:C}) can be rewritten as
\begin{equation}
\left \langle v^{2}(\mathbf{x}) \right \rangle=\sum_{\mathbf{k}}\left \langle \left | \tilde{v}_{\mathbf{k}} \right |^{2} \right \rangle=\sum_{j}\left [ \sum_{\mathbf{k}\in A_{j}} \left \langle \left | \tilde{v} _{\mathbf{k}} \right |^{2} \right \rangle \right ]=\sum_{j} S(k_{j}) \Delta K_{j} \simeq \int S(k)\; dk
\end{equation}
where we have introduced a new term $S(k_{j})$, which is the angle-integrated spectral density defined as
\begin{equation}
S(k_{j}) =\frac{1}{\Delta K_{j}}\sum_{\mathbf{k}\in A_{j}}\left \langle \left | \tilde{v}_{\mathbf{k}} \right |^{2} \right \rangle.
\end{equation}
The quantity $S(k_{j})$ is therefore related to an average over each radius,
\begin{equation}
S(k_{j})  =\frac{N_{j}}{\Delta K_{j}}\left \langle \left \langle \left | \tilde{v}_{k} \right |^{2} \right \rangle \right \rangle_{j} \simeq \frac{2\pi k_{j}}{\Delta k_{x} \Delta k_{y}}\left \langle \left \langle \left | \tilde{v}_{k} \right |^{2} \right \rangle \right \rangle_{j}=2\pi k_{j}\left \langle \left \langle  P(\mathbf{k}) \right \rangle \right \rangle_{j}.
\label{eq:D}
\end{equation}
The second expression comes from equating the annulus area $(A_{j}=2\pi k_{j} \Delta K_{j})$ to the area of each grid box $(\Delta k_{x}\Delta k_{y})$ in the definition of $P(\mathbf{k})$ in Equation (\ref{P:defined}), and by using the notation
\begin{equation}
\left \langle \left \langle \left | \tilde{v}_{k} \right |^{2} \right \rangle \right \rangle_{j} = \frac{1}{N_{j}} \sum_{\mathbf{k} \in A_{j}} \left | \tilde{v}_{\mathbf{k}} \right |^{2}
\end{equation}
to indicate the average in each annulus. Using the average here has the added benefit of producing an $S(k_{j})$ that is independent of window size.\par

All of the PSD measurements reported in this study use the angle-integrated spectral density given by Equation (\ref{eq:D}). \textit{It should be noted that if the quantity $S(k_{j})$ were to be calculated for a white noise source using the definitions given above, the resulting slope would have a value of +1 instead of the more customary flat spectrum. So the reader may wish to reduce by unity all the indices reported here when comparing $S(k_{j})$ to PSDs reported elsewhere. }

\section{Results}
\label{results_described}
This section presents the procedure by which the angle-integrated power spectral densities were calculated for intensity, kinetic energy, and enstrophy. It also details the information presented in Figures \ref{figure_intensity_PSD} -- \ref{psd_wrt_figure}, which are associated with these quantities.

\subsection{Intensity PSDs}
\label{Intensity_PSD_section}
All of the intensity images were first multiplied by a Hanning windowing function. This step is employed to mitigate any artifacts from edge effects due to the application of a Fourier transform to a function which is not actually periodic in space. The Hanning function is a favorable choice because of its spectral leakage characteristics, which provides a fair balance between spectral resolution and sensitivity to signal with low strength. Then the quantity
\begin{equation}
S_{I}(k_{j})   \simeq \frac{2\pi k_{j}}{\Delta k_{x} \Delta k_{y}}\left \langle \left \langle \left | \tilde{I}_{k} \right |^{2} \right \rangle \right \rangle_{j}
\end{equation}
was calculated for each frame, where $S_{I}(k_{j})$ is the angle-integrated intensity and $\tilde{I}_{k}$ is the Fourier transform of the contrast-enhanced scalar-field intensities. $\tilde{I}_k$ is the convolution of the underlying intensity PSD with $\tilde{W}_k$, the Fourier transform of the windowing functions, $W({\bf x})$. Therefore, the Fourier transform must be normalized to the windowing function by dividing the resulting $S_{I}(k_{j})$ by $\left \langle W^{2} \right \rangle$ to preserve the correct total integral. A similar normalization procedure was done for all of the angle-integrated spectral densities reported in this study.\par

After producing $S_{I}(k_{j})$ for each frame, a temporal average, $\left \langle S_{I}(k_{j}) \right \rangle$, was created by taking the geometric mean of $S_{I}(k_{j})$ in each k-bin. The geometric mean was used instead of the arithmetic mean because the resulting values of $S_{I}(k_{j})$ had a variance covering several orders of magnitude in each bin. The intensity values were initially in units of DN s$^{-1}$, but are now in arbitrary units after rescaling the images with our contrast enhancement method. An example of $\left \langle S_{I}(k_{j}) \right \rangle$ is shown in Figure \ref{figure_intensity_PSD} with two distinct power-law regions, indicated by the vertical blue and green dashed lines, fitted by a least-squares method. The corresponding power-law fits are indicated by the green and blue dash-dot lines. The dash-vertical red lines highlight two of the most predominant spikes in the spectra due to the JPEG compression used by SOT. The other peaks correspond to additional harmonics related to this compression. The range of wavenumbers used for performing the least-squares fit and the resulting indices calculated for the two power-laws in each intensity spectrum can be found on Table \ref{table:C}. The uncertainty ($\sigma_{fit}$) in the fits was estimated by
\begin{equation}
\sigma_{fit} =\sqrt{\frac{\chi ^{2} }{\nu }}
\label{uncertain_eq}
\end{equation}
where $\chi ^{2}$ is the unreduced chi-square and $\nu$ is the number of degrees of freedom. This is the formulation used for all the uncertainties reported in the indices shown in Table \ref{table:C}.

\subsection{Kinetic Energy and Enstrophy PSDs}
\label{Kinetic Energy and Enstrophy PSDs}
The two-dimensional velocity field within the prominences can be denoted as $\mathbf{v}(\mathbf{x})=u(\mathbf{x})\mathbf{\hat{x}}+v(\mathbf{x})\mathbf{\hat{y}}$. The corresponding discrete Fourier transform of each component is denoted by $\tilde{u}_{k}$ and $\tilde{v}_{k}$. Now the angle-integrated kinetic energy spectrum $S_{KE}(k_{j})$ can be determined by
\begin{equation}
S_{KE}(k_{j})=\frac{1}{\Delta K_{j}} \sum_{\mathbf{k} \in A_{j}} \left (  \frac{1}{2} \left | \tilde{u}_{\mathbf{k}} \right |^{2} +\frac{1}{2} \left | \tilde{v}_{\mathbf{k}} \right |^{2} \right )
\label{Equation_of_SKE}
\end{equation}
and the average kinetic energy density per unit mass ($\varepsilon$) can then be determined from
\begin{equation}
\varepsilon =\left \langle \frac{1}{2}\left | \mathbf{v}\right |^{2} \right \rangle=\sum_{j}S_{KE}(k_{j})\Delta K_{j}.
\label{total_energy_eq}
\end{equation}

Several eddies can also be seen in these prominences at varying length scales (see movies available with the online version). One way to quantify how these eddies transfer kinetic energy to the smaller length scales -- eventually to the dissipation scale -- is by measuring the system's enstrophy PSD. First, the vorticity of the two-dimensional velocity field can be denoted as $\zeta (\mathbf{x})=\mathbf{\hat{z}}\cdot (\triangledown \times \mathbf{v})$. Enstrophy can then be expressed as the square modulus of $(\triangledown \times \mathbf{v})$. But direct calculation of $(\triangledown \times \mathbf{v})$ introduces a lot of noise due to large gradients in the velocities, especially at small length scales. Since it is the PSD we want in the end, and not the enstrophy per se, we can take advantage of calculating in the Fourier domain to avoid the worst of the noise issues. That is, the square modulus of the curl is tantamount to multiplication by the square of the wavenumber. Thus, the angle-integrated enstrophy spectrum $S_{EN}(k_{j})$ can be determined by
\begin{equation}
S_{EN}(k_{j})=\frac{1}{\Delta K_{j}}\sum_{\mathbf{k} \in A_{j}}\left | \tilde{\zeta} _{\mathbf{k}} \right |^{2}=\frac{1}{\Delta K_{j}}\sum_{\mathbf{k} \in A_{j}}\left [ k^{2}_{y}\left |  \tilde{u}_{\mathbf{k}}\right |^2+k^{2}_{x}\left |  \tilde{v}_{\mathbf{k}}\right |^2-2k_{x}k_{y}\mathrm{Re}(\tilde{u}_{\mathbf{k}}\tilde{v}^{*}_{\mathbf{k}})\right ]
\label{enstrophy_eq}
\end{equation}
where $\tilde{\zeta}_{\mathbf{k}}=ik_{x}\tilde{v}_{\mathbf{k}}-ik_{y}\tilde{u}_{\mathbf{k}}$ is the Fourier transform of $\zeta (\mathbf{x})$. The average enstrophy density per unit mass ($\omega$) can be calculated from
\begin{equation}
\omega =\left \langle \frac{1}{2} \left | \zeta(\mathbf{x}) \right |^{2}\right \rangle=\sum_{j}\frac{1}{2}S_{EN }(k_{j})\Delta K_{j}.
\label{total_enstrophy_eq}
\end{equation}

The velocity maps were prepared by zeroing out values at the outer border and then multiplying the results by a Hanning function. $\left \langle S_{KE}(k_{j}) \right \rangle$ and  $\left \langle S_{EN}(k_{j}) \right \rangle$ were both calculated in the same way as $\left \langle S_{I}(k_{j}) \right \rangle$ in Section \ref{Intensity_PSD_section}. An example of  $\left \langle S_{KE}(k_{j}) \right \rangle$ and $\left \langle S_{EN}(k_{j}) \right \rangle$ can be seen for the first set of even-numbered frames for the 2006/11/30 prominence in Figure \ref{figure_KE_and_enstrophy_PSD}. The red dash-dot line indicates the linear fit and the red solid vertical lines indicate the corresponding fit range. The uncertainty in the fit parameters were determined by Equation (\ref{uncertain_eq}). A complete list of the indices, uncertainties, and fit ranges are all indicated in Table \ref{table:C}. The red vertical error bars represent 3 geometric standard errors of the geometric mean associated with each $\left \langle S_{KE}(k_{j}) \right \rangle$ and  $\left \langle S_{EN}(k_{j}) \right \rangle$ data point, but only the region to the left of the fit range has the uncertainties shown to avoid cluttering the plot. This illustrates what the typical error bar size looks like. We only included pixels where a velocity could be determined at least 95\% of the time in our analysis of $S_{KE}(k_{j})$ and $S_{EN}(k_{j})$. This was easily implemented by using the Boolean map from the FLCT, which indicates if a velocity was calculated at a given pixel location.\par

An additional test was performed to determine the effect all the FLCT parameters had on the derived velocity fields. First, a synthetic image was created by specifying a desired power law structure in k-space and then some random phase information was added for populating these powers in the spatial-frequency space. The inverse Fourier transform of this array was then used as the initial image. This procedure was then repeated to create two arrays of synthetic velocities $u(x,y)$ and $v(x,y)$ with power law spectra specified independently from that of the intensity. These velocities were used to advect the original intensity image, yielding a second image $\delta t$ later in time. The two synthetic images were then used as the input to the FLCT program and the spectra of these inferred velocity fields were compared to the power law specified in the generation of the actual fields. \changes{(An example of this technique can be found in Section \ref{Appendix}.)}\par

This exercise showed that $\sigma$ (the apodization window) had the greatest effect on the calculated velocity power spectral density, which was also stated in similar test performed by \cite{Welsch04}. The results found with the FLCT program started to depart from the known velocity in wavenumber-space around values corresponding to a wavelength of 1$\sigma$ to 2$\sigma$. Therefore, any results in the velocity power spectral density with a wavelength less than 2$\sigma$ cannot be trusted and will not be included in any of our analysis. This test also confirmed that the parameters chosen by visually inspecting the corks in the derived field were the best combination, but this test gave a more quantitative reason for choosing them.\par

The two blue vertical lines in Figures \ref{figure_KE_and_enstrophy_PSD} correspond to wavelengths related to the FLCT apodization window size. The test described above has shown that powers in the velocity spectrum to the right of these blue lines are unreliable. Also, $S_{KE}(k_{j})$ was calculated in three distinct ways, by using either the horizontal, vertical, or both velocities in Equation (\ref{Equation_of_SKE}). These results can be found in Table \ref{table:C} and an example of the difference is emphasized in the bottom of Figure \ref{PSDs_k_and_velocity_components} for the first even-numbered data set of the 2006/11/30 prominence. The temporal evolution of the indices produced by $S_{KE}(k_{j})$ and $S_{EN}(k_{j})$ are given in Figure \ref{psd_wrt_figure}. The indices shown by the black line have been smoothed by a boxcar with a width of 3 time steps. The indices from $\left \langle S_{KE}(k_{j}) \right \rangle$ are also shown for results computed from the horizontal ($u(\mathbf{x})$), vertical ($v(\mathbf{x})$), or both velocities. The horizontal error bars indicate the time span used for calculating the temporally smoothed $\left \langle S_{KE}(k_{j}) \right \rangle$ and $\left \langle S_{EN}(k_{j}) \right \rangle$ results.  All of the indices from kinetic energy, enstrophy, and intensity spectra at low wavenumbers were fitted using the same wavenumber range to make comparison easier. Discussion of the spectra and trends observed in the power-law indices is found in the next section.

\section{Discussion}
\label{Discussion_section}

The spectra of intensities, enstrophy, and kinetic energy tell us about the distribution of these characteristics inside the prominence at different length scales. It is hoped that analyses of these properties -- their PSD indices, and breaks and anisotropies in the same -- will contribute to a better understanding of the complex relationship between the mass motions and the embedded magnetic field, potentially including the tangling and distortion of the prominence magnetic field \citep[cf.][]{Ball08}. The indices found from the fitted power laws of the spectra are used as a proxy for describing how these characteristics vary over a range of wavenumbers. In this section, we consider some aspects that are evident in the spectra reported here. We shall adopt the convention of referring as `harder' to spectra with relatively more power in the high-wavenumber region, while those with less high-k power are called `softer'.\par

There is a noticeable break in the PSD of image intensity $\left \langle S_{I}(k_{j}) \right \rangle$ around $k\approx 3-5 \; \textrm{rads Mm}^{-1}$ as seen in Figure \ref{figure_intensity_PSD}. This is reminiscent of a similar break seen in Figure 3 of \cite{Leonardis12} around $k\approx 2-3 \; \textrm{rads Mm}^{-1}$, which was found by performing a comparable analysis but with one-dimensional strips in the horizontal and vertical directions for the same prominence. However, the spectra we observed by analyzing the entire two-dimensional image were considerably softer than those reported by \cite{Leonardis12} for both the low and high wavenumber regions. It is unclear if this break is indicating a change in the physics taking place at different length scales or merely a noise component contributing to the spectrum at small wavenumbers. We note that \cite{Leonardis12} performed their analysis apparently after contrast-enhancement of the image intensities, as did we.  To investigate whether the break in the spectral index might be an artifact of the contrast enhancement, we applied the identical PSD analysis to the un-enhanced SOT images.  It was determined that although the slopes in the two power-law regions changed when no contrast enhancement was applied (for example the slopes in Figure \ref{figure_intensity_PSD}  went from -1.63 to -1.52 and from -2.25 to -2.30), the break was unchanged: the PSD still shows a break in the power-law index at the same wavenumber as with the contrast enhancement. The indices reported here for $\left \langle S_{I}(k_{j}) \right \rangle$ and $\left \langle S_{KE}(k_{j}) \right \rangle$ are also significantly different, which implies that the intensity and velocity structures have different scaling relationships. There is also a break in the spectra of $\left \langle S_{KE}(k_{j}) \right \rangle$ and $\left \langle S_{EN}(k_{j}) \right \rangle$, but it is too close to the region where the FLCT has difficulty producing credible results. Hence it cannot be stated conclusively if the knees in these power laws are real or an artifact, but we think it very unlikely that the dissipation range extends to length scales as large as 2 Mm.\par 

The kinetic energy and enstrophy spectra tell us that there is self-similarity present in these structures, which is evident by the observed power law. The kinetic energy power spectral density, $\left \langle S_{KE}(k_{j}) \right \rangle$, produced indices ranging from 0.00 to -0.60 (below the +1 expected for white noise) and the enstrophy spectral density $\left \langle S_{EN}(k_{j}) \right \rangle$ showed indices between 1.11 and 1.65.  Figure \ref{figure_KE_and_enstrophy_PSD} suggests a cascade to smaller length scales in the interval of $0.8 \leq k\leq (2.0-3.2)\; \textrm{rads} \: \textrm{Mm}^{-1}$ for kinetic energy and the spectra of $\left \langle S_{KE}(k_{j}) \right \rangle$ is also considerably softer than $\left \langle S_{EN}(k_{j}) \right \rangle$. The kinetic energy PSDs shown in Figure \ref{psd_wrt_figure} have two horizontal dotted lines that indicate the indices corresponding to Iroshnikov-Kraichnan's $(k^{-3/2})$ and Kolmogorov $(k^{-5/3})$ turbulence models \citep{Kraichnan65,Kolmogorov91} in terms of angle-integrated spectral density values. The majority of the time the indices are less steep than these values, but there are a few instances where they could be displaying behavior that correspond to one of the before mentioned turbulence theories. We do not expect the flows within the prominences to be freely decaying turbulence like that described by Kolmogorov or Iroshnikov-Kraichnan. Instead it is driven by gravitational and magnetic pressure within the prominence. The self-similarity we observed suggests the driving is occurring at length scales greater than 7.9 Mm.\par

The positive slope in enstrophy PSD, slightly higher than would be expected for white noise, indicates the presence of increased angular velocities at smaller length scales due to, e.g., generation of new vortices at smaller length scales, or vortex stretching, or a combination thereof.  However, it is difficult to be more precise about the nature of the enhanced vorticity without three-dimensional velocity measurements.

The index values of $S_{KE}(k_{j})$ and $\left \langle S_{KE}(k_{j}) \right \rangle$ are found to be below -0.5, which is similar to results found in solar wind measurements \citep{Boldyrev11}. In the case of solar wind, the spectra from the magnetic field energy are slightly steeper than -0.5 and kinetic energy spectra are harder, thereby producing an index that is slightly less than -0.5. However, the spectral index produced from the sum of these two energies is $k^{-3/2}$ \citep{Boldyrev11}, which corresponds to strong MHD turbulence. One could conjecture that a similar process is occurring with the spectra of these quiescent prominences. However, we still need more information on the energy of magnetic fields residing inside these prominence; and must concede that the $\beta$ values that dictate the solar wind's behavior are considerably different than in prominences. But, magnetic field measurements made with THEMIS \citep{Schmieder14} show a promising indication of being able to map the magnetic field and velocities of the total prominence in the near future.\par 

Another possible explanation for the observed index values is that the system has not yet reached a fully developed turbulent state. There is some indication of this being the case when examining the temporal evolution of the 2007/10/03 prominence in Figure \ref{psd_wrt_figure}. This prominence begins with a rather large cavity that erupts away from the solar limb and then it begins to reach a more stationary or equilibrium state (see intensity movie of 2007/10/03 prominence). The initial configuration starts with the index of $\left \langle S_{KE}(k_{j}) \right \rangle$ at approximately -0.2, but it then steadies out around -0.5 after the large cavity is evacuated. This episode also shows how the $S_{KE}(k_{j})$ spectrum responds to a large driving force. \par

Figures \ref{PSDs_k_and_velocity_components} \& \ref{psd_wrt_figure}  illustrate how the slopes calculated for $\left \langle S_{KE}(k_{j}) \right \rangle$ can be altered when using the Fourier transform of the horizontal ($u$), vertical ($v$), or both velocities in Equation (\ref{Equation_of_SKE}). The results clearly show an anisotropy of the flows. The vertical velocity component, $v$, has a notably harder spectrum than the horizontal component, $u$. The vertical motion seen in the intensity movies are the result of upward flows originating at the convection bubbles near the base of the prominences, which compete with gravity driven flows moving back down to the solar surface. The horizontal flows, on the other hand, seem to result from the deflection of the vertical flow by a tilted magnetic field.  In this sense the horizontal flow is {\em driven} by the vertical flow. All of these competing motions are superimposed onto each other, which results in vortices and the observed anisotropic behavior of $\left \langle S_{KE}(k_{j}) \right \rangle$. We plan to conduct a followup investigation on how the horizontal velocities might be altered by projection effects near the solar limb.  \par 

There are also two of every symbol listed in the legend of the plots shown in Figure \ref{psd_wrt_figure} for the 2006/11/30 and 2007/04/25 prominence. These identical data points for a given time range corresponds to the results calculated from using the even and odd frames. Most of the symbols cannot be distinguished and must be examined in Table \ref{table:C} to tell the difference. The even and odd frame sets should have values close to each other since the measurements were made over the same timespan. Therefore, this can be used as another indication of the magnitude of the error associated with these measurements. \par

Separate insight into the anisotropy can be gained from examining one-dimensional cuts through the two dimensional PSD $\langle P(k_x,k_y) \rangle$. The top plot of Figure \ref{PSDs_k_and_velocity_components} shows three different results for calculating the temporal averaged power spectral density, $\left \langle P_{KE}(k) \right \rangle$, for the last even data set of the 2006/11/30 prominence. The green asterisks and blue diamonds show horizontal and vertical cuts, respectively, 
$\langle P(k_x,0) \rangle$ and $\langle P(0,k_y) \rangle$. These values are compared to $\left \langle P_{KE}(k) \right \rangle$ calculated with $k=\sqrt{k_{x}^2+k_{y}^{2}}$, i.e., the magnitude of the radial wave-vector. This illustrates that there is an anisotropic nature to the velocities in Fourier space and they correspond to the behavior seen when examining $\left \langle S_{KE}(k) \right \rangle$ at the bottom of Figure \ref{PSDs_k_and_velocity_components}. \par

It is instructive to also examine how the spectra might vary as a function of spatial location. This was done by taking a closer look at different regions associated with the 2006/11/30 prominence. We chose this prominence because it had a noticeable amount of plasma in the foreground and some spicule motion at the bottom, which can affect the calculated spectra. It also had more distinct regions with varying characteristics when compared to the other two prominences. Three Regions Of Interest (ROI) were selected and are shown in Figure \ref{intensity_figure}. The motions in ROI 1 consist of small plumes rising into the field of view with the occasional large curling motion in the plasma sheet. ROI 2 is located above a large convection bubble at the base of the prominence and a large plume originating from the bubble is caught entering the field of view around 06:30 UT. ROI 3 was chosen because it has a noticeable foreground component in the images over the entire time span that resembles a possible overlapping barb structure (see \cite{Vial15} for description of filament barbs). This overlap is also evident via a prominent double peak in the intensity value distributions, which shows the contribution from the foreground material is significantly brighter. By examining ROI 3, we can determine how the results will vary when the two-dimensional assumption used in this work breaks down.\par

The same kinetic energy and enstrophy analyses described in Section \ref{results_described} were conducted on these three 128$\times$128 pixel ROIs for the last even data set of the 2006/11/30 prominence. These results were boxcar smoothed with a width of 3 time steps and presented in Figure \ref{ROI_KE_enstrophy} for comparison. The mean and standard deviation of the indices produced can also be found in Table \ref{ROI_results}. There is a time period where a large plume enters the field of view of ROI 2 and is indicated in Figure \ref{ROI_KE_enstrophy} by the red solid vertical lines. During this time interval there is noticeable softening in $S_{KE}$ and $S_{EN}$ spectrum for ROI 2 as compared to the other regions and the total Field Of View (FOV). At the same time the kinetic energy density is shown to increase for ROI 2 in Figure \ref{ROI_energy_plot}, as described in detail below. Table \ref{ROI_results} also shows that ROI 2 has the steepest slopes associated with $\left \langle S_{KE}(k_{j}) \right \rangle$ and also experiences the greatest anisotropy in indices for $\left \langle S_{KE}(k_{j}) \right \rangle$. This makes sense since the large rising plumes will contribute greater to the vertical than the horizontal velocities. ROI 1 and 3 both have less negative slopes when compared to the total FOV, which illustrates that these regions contribute to a flattening of the overall-FOV spectrum. There is a slight difference in the characteristics of ROI 3, which suggest that the effect of foreground plasma in the images cannot be ruled out as contributing to results for the entire FOV. It should be noted that foreground material is present but hard to distinguish in the H-alpha images of the 2007/04/25 prominence due to its resolution and it is believed to be a major factor in the large variations in its index values shown in Figure \ref{psd_wrt_figure}. However, there is no such contribution seen in the 2007/10/03 prominence. \par

\begin{figure}
\centering
   \includegraphics[angle=90,scale=0.7]{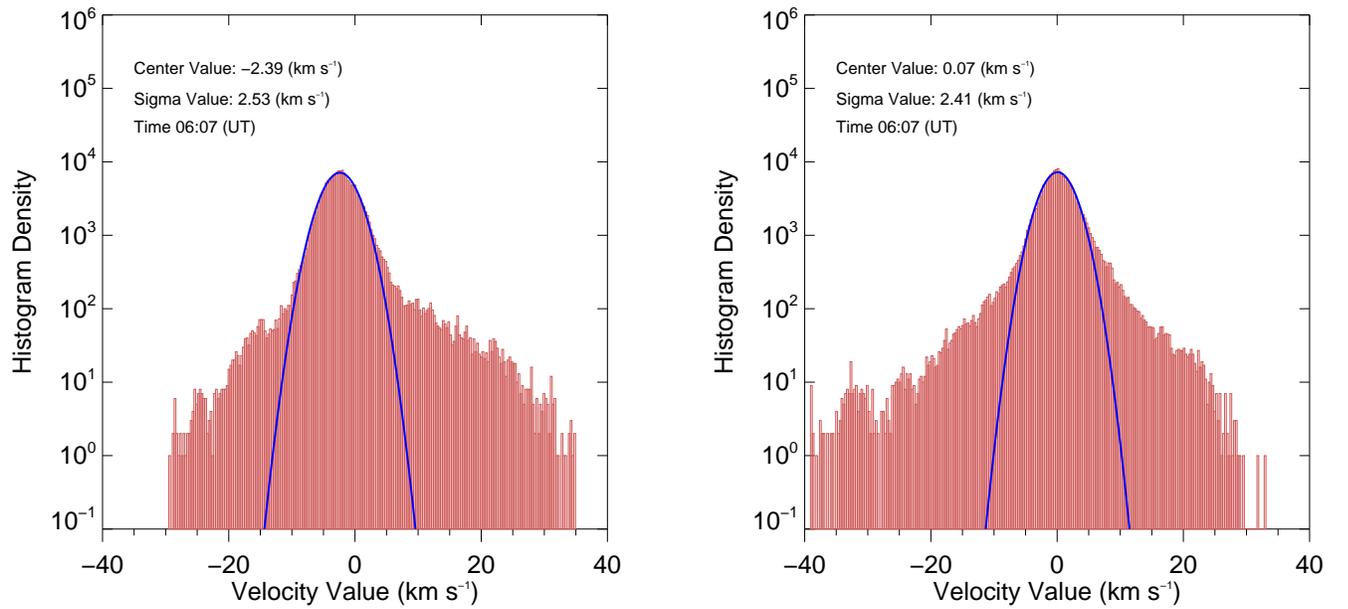}\\
  \caption{These histograms show the horizontal (left) and vertical (right) velocity distributions for the 2006/11/30 prominence at 06:07:50 UT. The lines overlaid on top indicates a Gaussian fit applied to the binned data with the center/mean and $\tilde{\sigma}$ given in the top left corners. Each bin has a width of 0.33 km s$^{-1}$.}
  \label{histogram_of_velocities}
\end{figure}

\begin{figure}
\centering
  \begin{tabular}{@{}c@{}}
   \includegraphics[angle=90,scale=0.40]{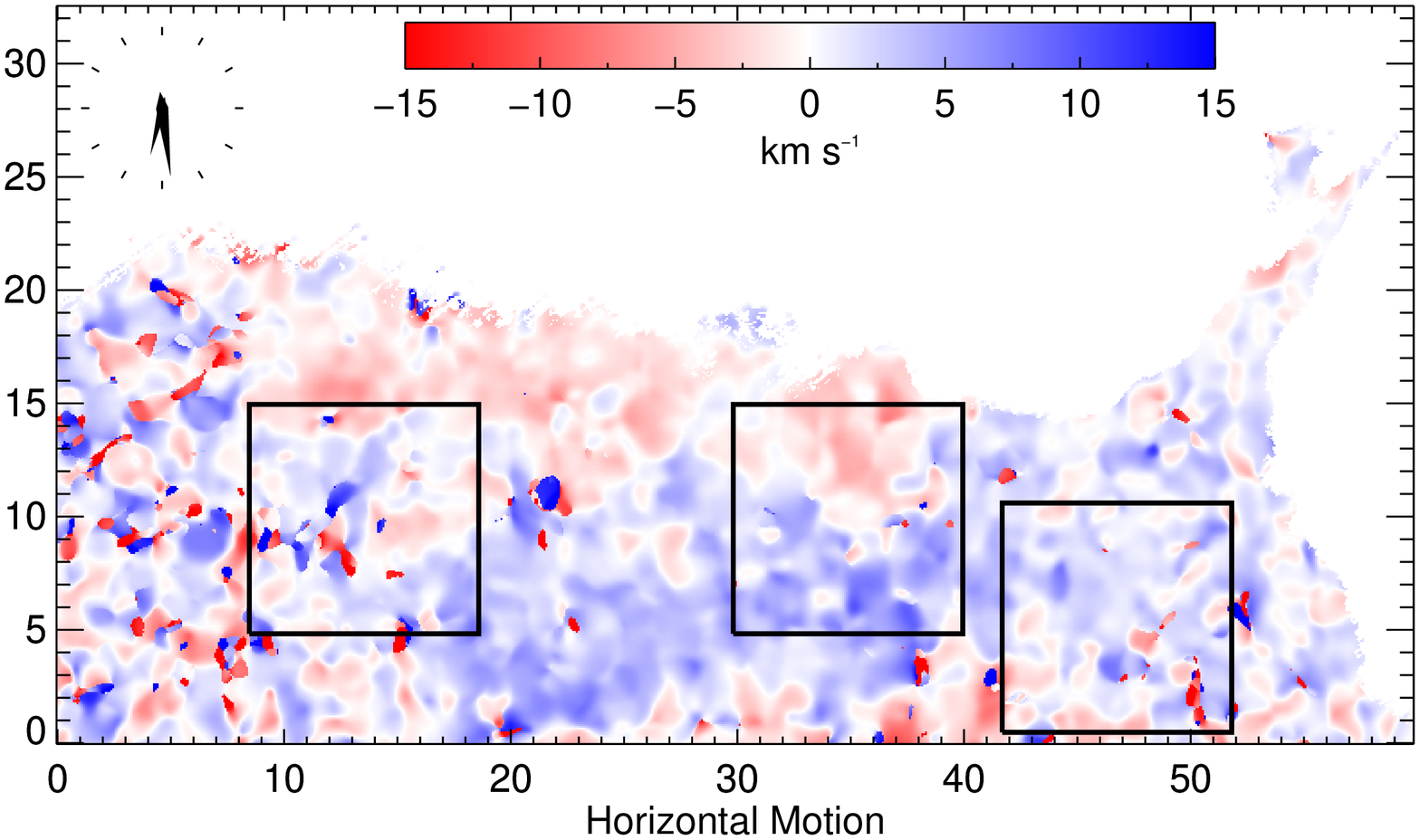}\\
   \includegraphics[angle=90,scale=0.40]{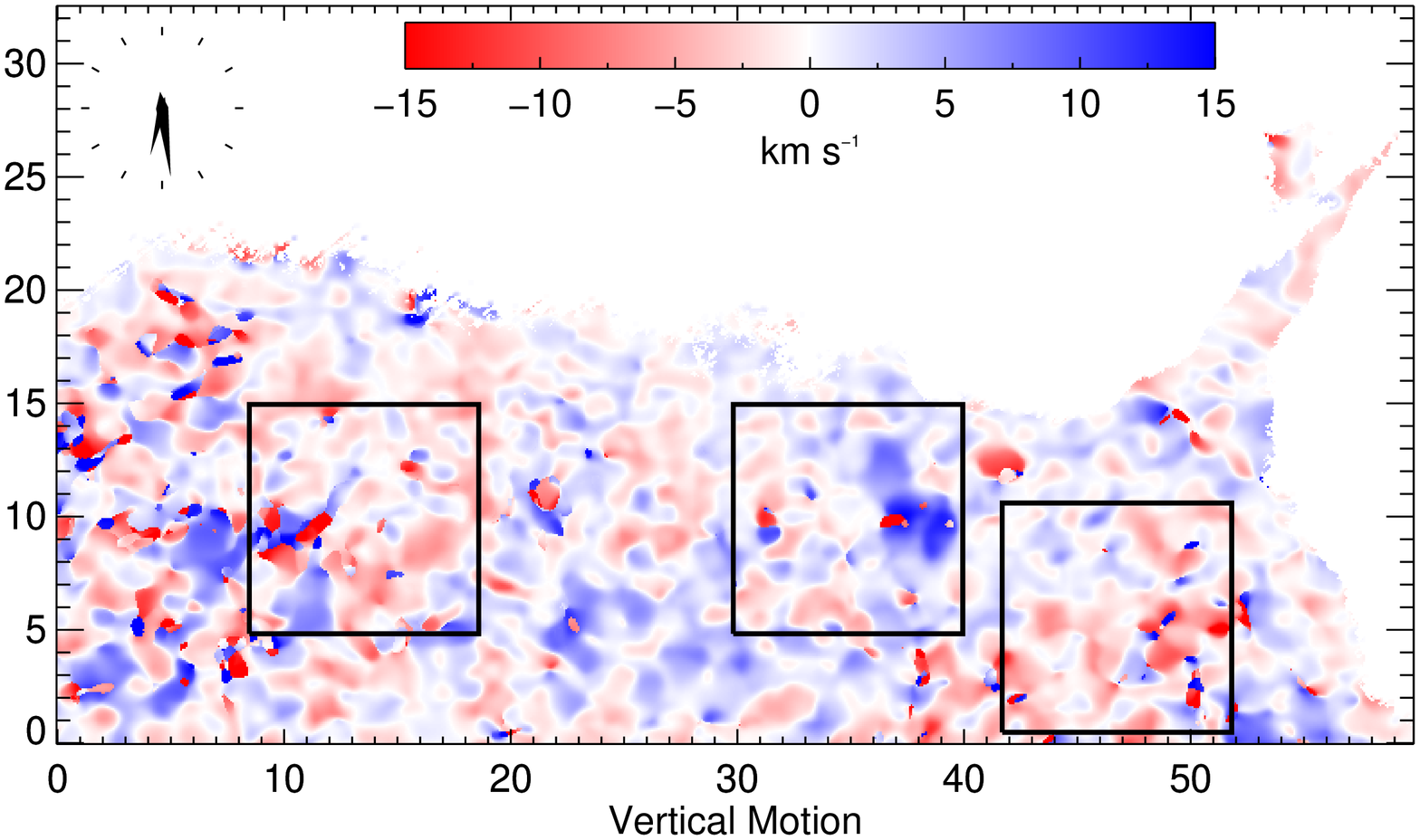}\\
  \end{tabular}
\caption{Example of velocity map for the intensity image of 2006/11/30 prominence shown in Figure \ref{intensity_figure}. Positive velocities in these figures indicate horizontal motion to the right and vertical motions moving away from the solar surface. A plume is moving upwards in ROI 2 during this time and exhibiting speeds similar to those reported by \cite{Berger10}. Movies showing all of the prominence velocities can be found online with the electronic version. Axis are given in units of Mm. Velocity values can exceed $\pm$ 15 km s$^{-1}$ as indicated by Figure \ref{histogram_of_velocities}, but they have been truncated here to make the dynamic range of velocities easier to illustrate. The boxes shown here correspond to ones from Figure \ref{intensity_figure}.}
\label{figure_of_velocity_map}
\end{figure}

\begin{figure}
\centerline{\includegraphics[scale=0.50]{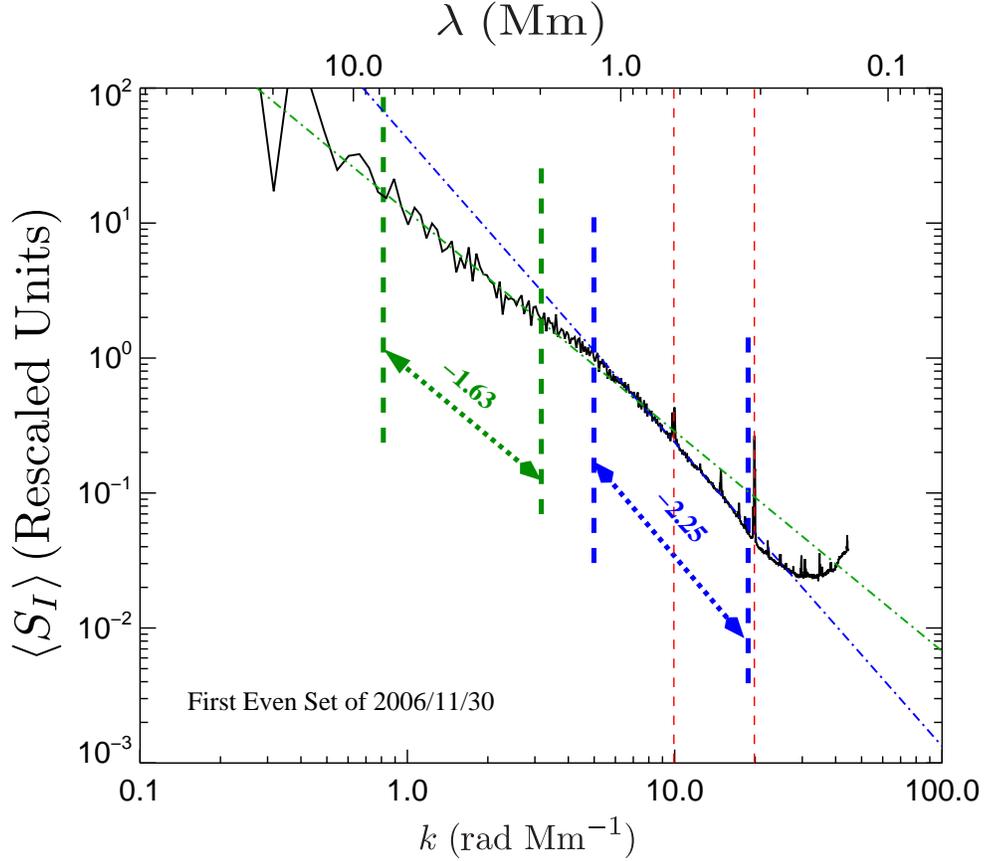}}
\caption{This is an example of $\left \langle S_{I}(k_{j}) \right \rangle$ for the first even data set associated with the prominence on 2006/11/30. The spikes in the spectrum are the result of the jpeg compression used on the SOT telemetry. The largest spikes are emphasized by the red vertical dashed lines, which indicate wavelengths corresponding to 8 and 4 pixels in magnitude. An index of $-2.25\pm0.02$ is shown for the fit range of $5<k<19$ rads Mm$^{-1}$, as indicated by the dashed blue vertical lines. A fit is also performed for the range of $0.8<k<3.2$ rads Mm$^{-1}$, which is indicated by the dashed green vertical lines and resulted in an index of $-1.63\pm0.07$. The green and blue dot-dash lines show the fits found for these two regions. The intensity values are in arbitrary values due to the contrast enhancement applied. }
\label{figure_intensity_PSD}
\end{figure}

\begin{figure}
\centering
  \begin{tabular}{@{}c@{}}
   \includegraphics[scale=0.35]{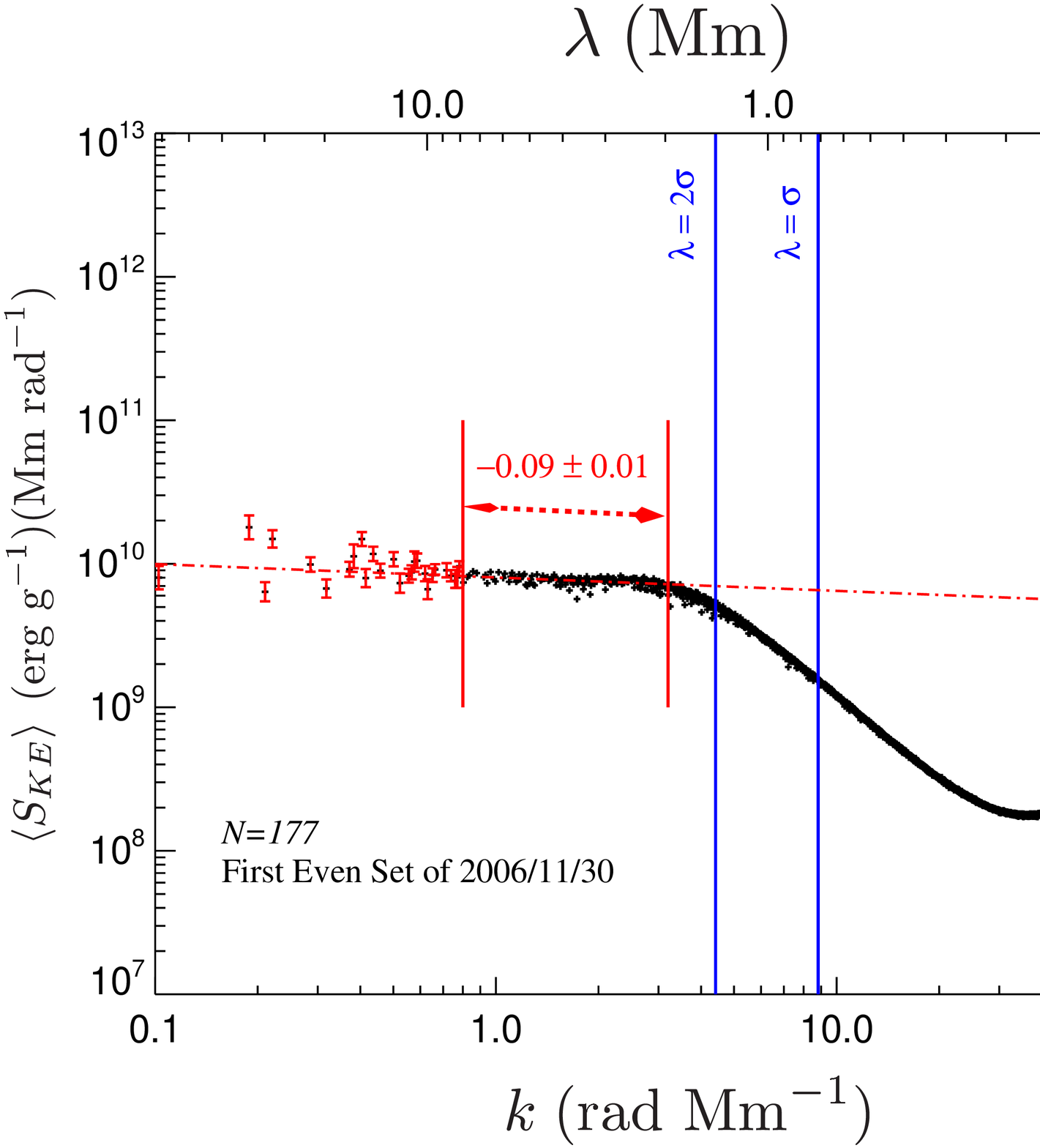}\\
   \includegraphics[scale=0.36]{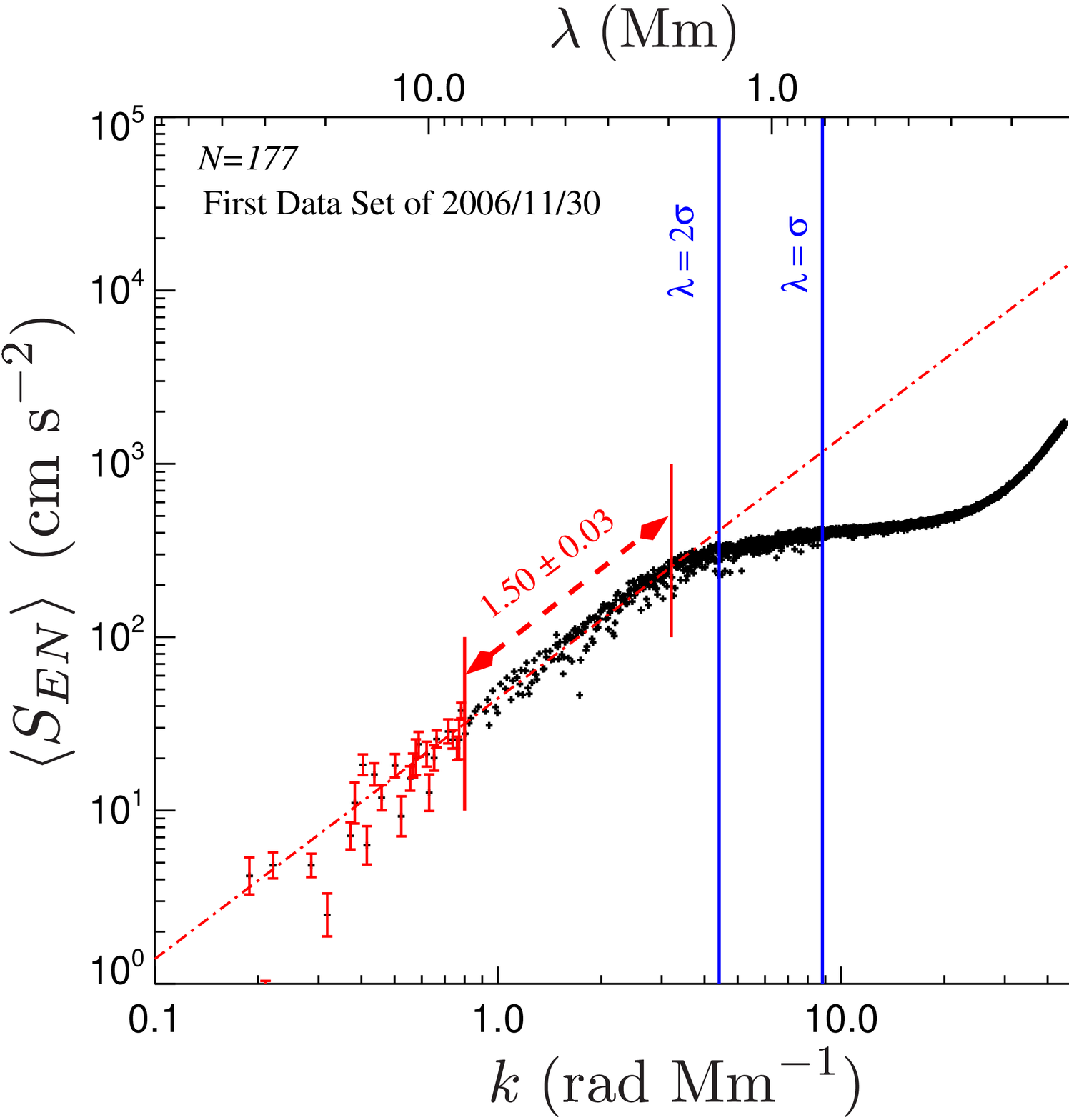}\\
  \end{tabular}
\caption{The typical $\left \langle S_{KE}(k_{j}) \right \rangle$ (top) and $\left \langle S_{EN}(k_{j}) \right \rangle$ (bottom) for the first even data set associated with the prominence on 2006/11/30. These plots corresponding to the filled black circle furthest to the left, in the top-left plot of Figure \ref{psd_wrt_figure}. Error bars are included at the low wavenumber values, for clarity, but they are similar in size at all wavenumbers. The spectra indices are determined by fitting between the red vertical lines. The velocity resolution from the FLCT is indicated by the blue vertical lines and therefore  $k\;>\; (\pi/\sigma)$ are excluded from our analysis.}
\label{figure_KE_and_enstrophy_PSD}
\end{figure}

\begin{figure}
\centering
  \begin{tabular}{@{}c@{}}
   \includegraphics[scale=0.34]{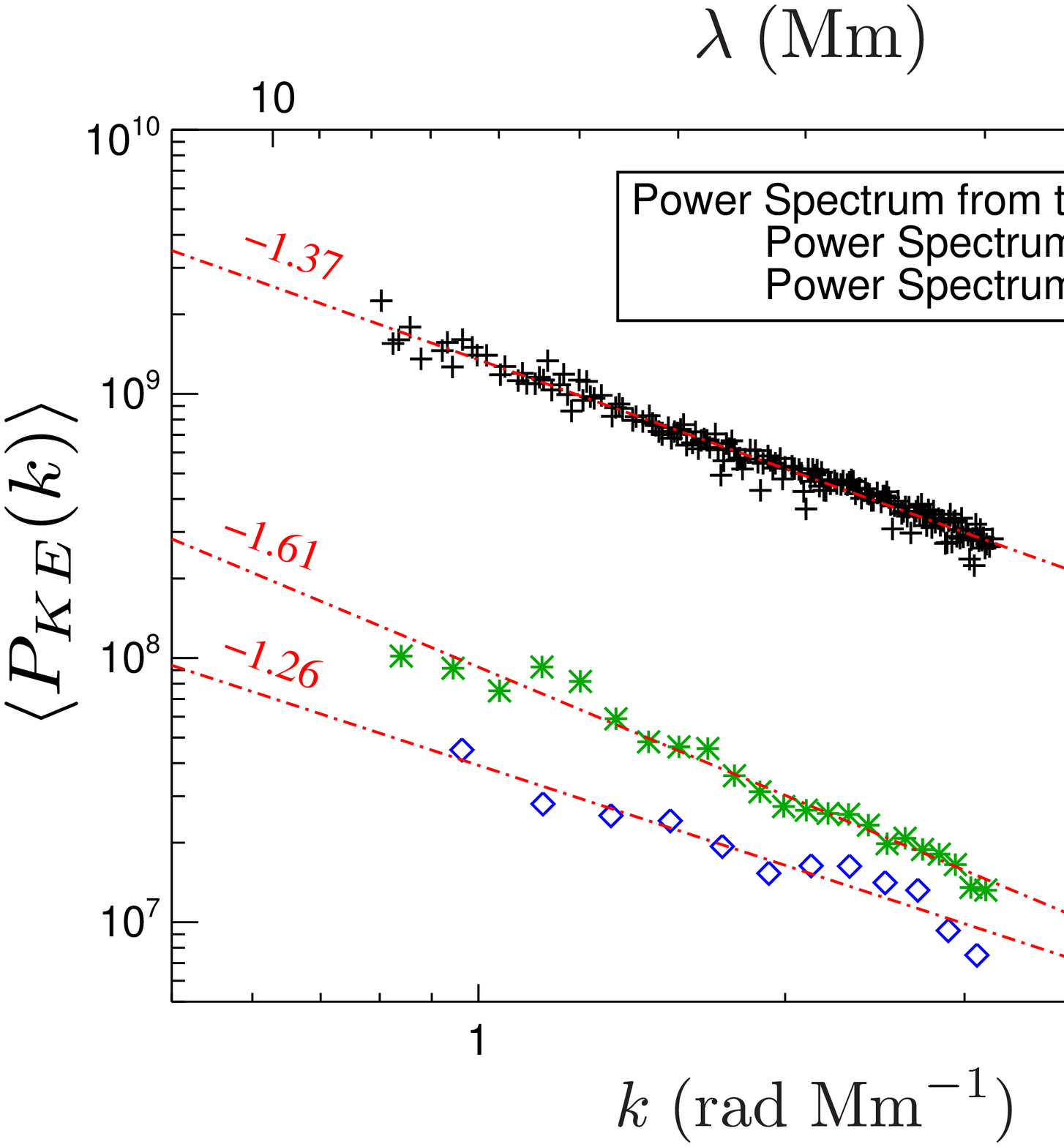}\\
   \includegraphics[scale=0.5]{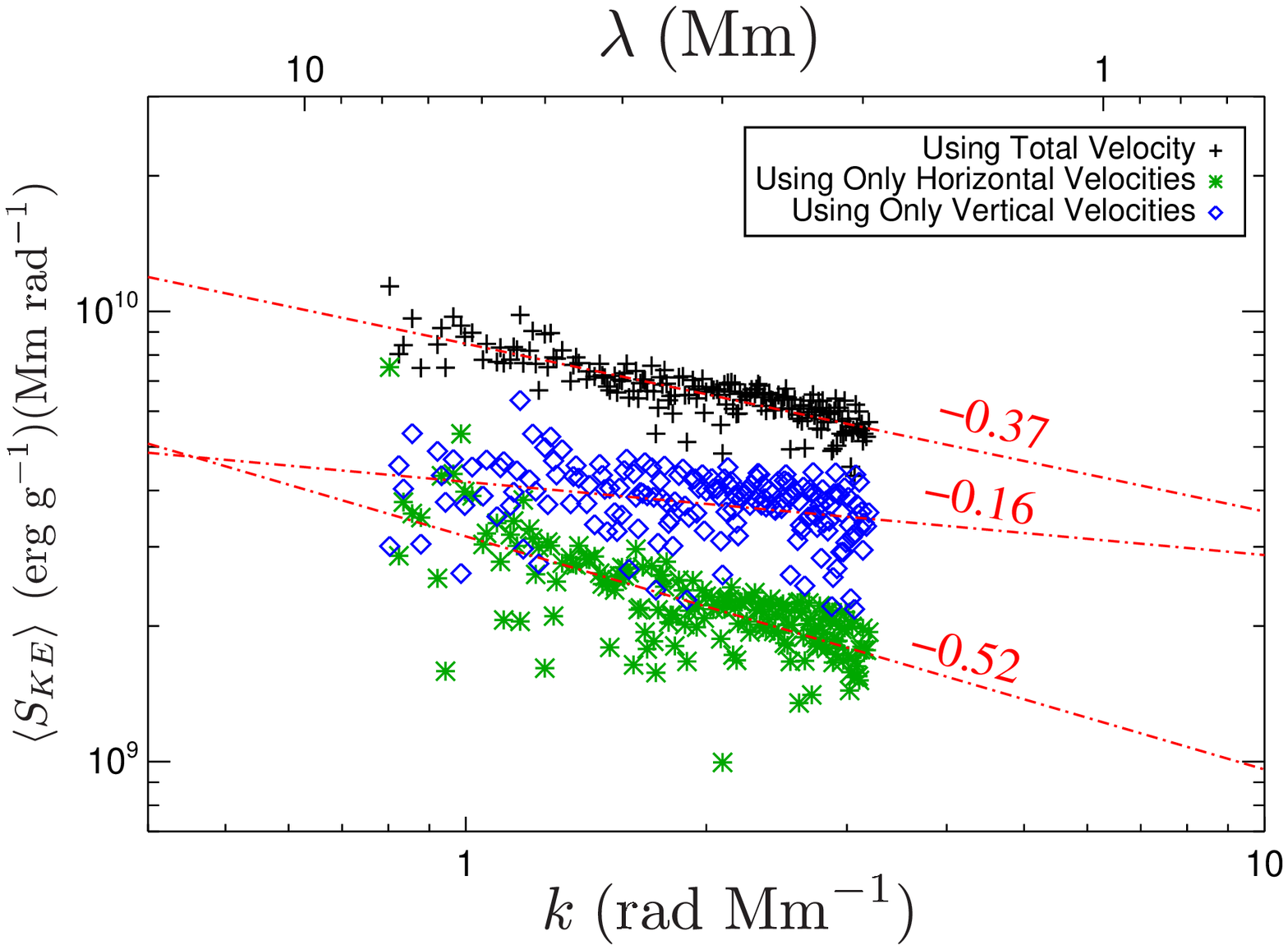}\\
  \end{tabular}
\caption{Both figures are from the last even data set of the 2006/11/30 prominence. The top figure shows the kinetic energy power spectrum density, $\left \langle P_{KE}(k) \right \rangle$, found by examining along the $\langle P(k_x,0) \rangle$ (green asterisk), $\langle P(0,k_y) \rangle$ (blue diamonds), and over the entire wavenumber space (black pluses). $\left \langle P_{KE}(k) \right \rangle$ has units of (erg s$^{-1}$)(Mm rad$^{-1}$) for the spectra along the k-axes and (erg s$^{-1}$)(Mm$^{2}$ rad$^{-2}$) when examining the entire Fourier space. This illustrated how anisotrophic the power distribution is in Fourier space and how our reported values compare. The bottom figure compares the kinetic energy PSD produced when only using horizontal (green asterisk), vertical (blue diamonds), or total/both velocities (black pluses) in Equation (\ref{Equation_of_SKE}). All the indices found from fitting the data is indicated above the associated fit line.}
\label{PSDs_k_and_velocity_components}
\end{figure}

\begin{figure}
\centering
  \begin{tabular}{@{}cc@{}}
   \includegraphics[angle=90,scale=0.34]{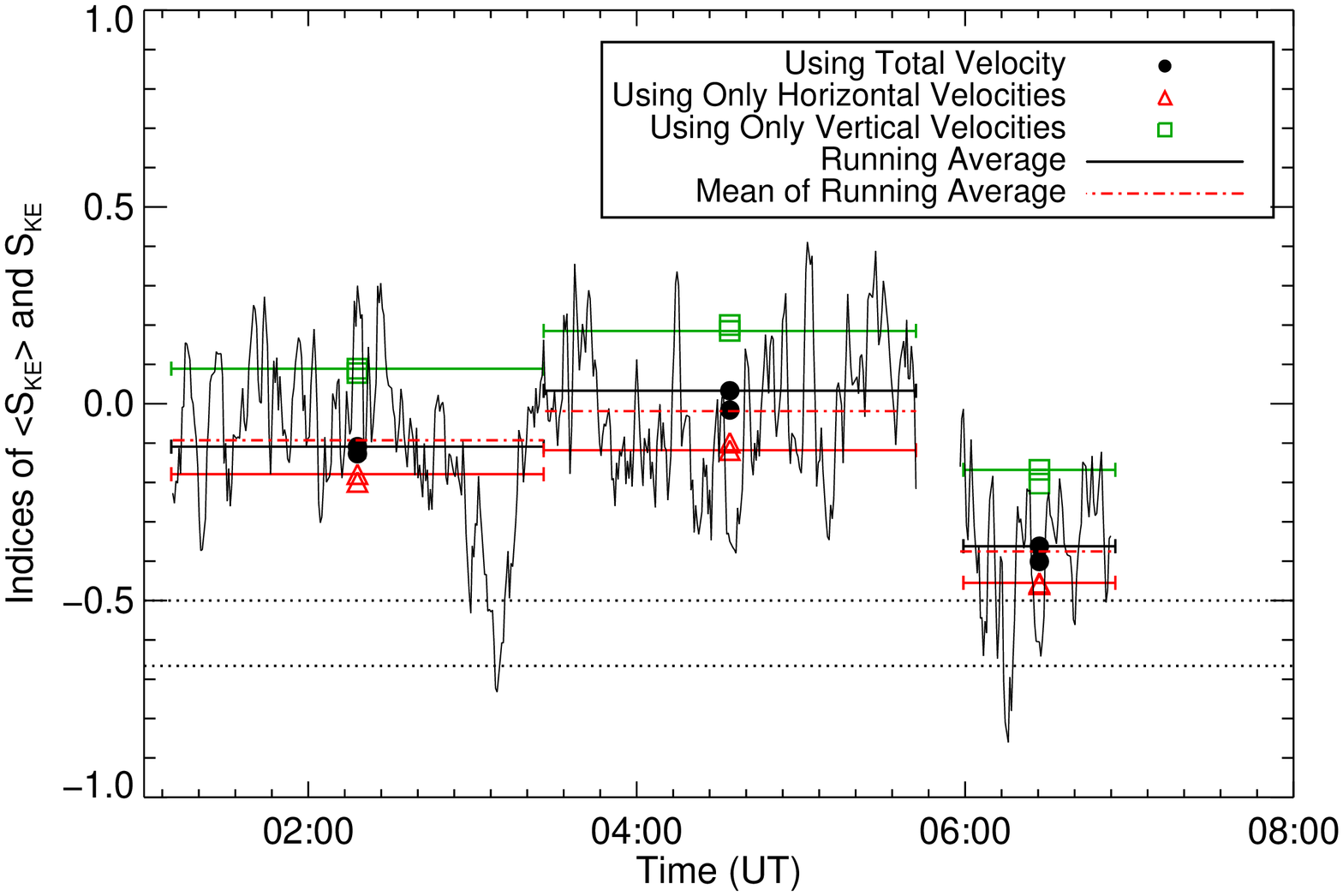} & \includegraphics[angle=90,scale=0.34]{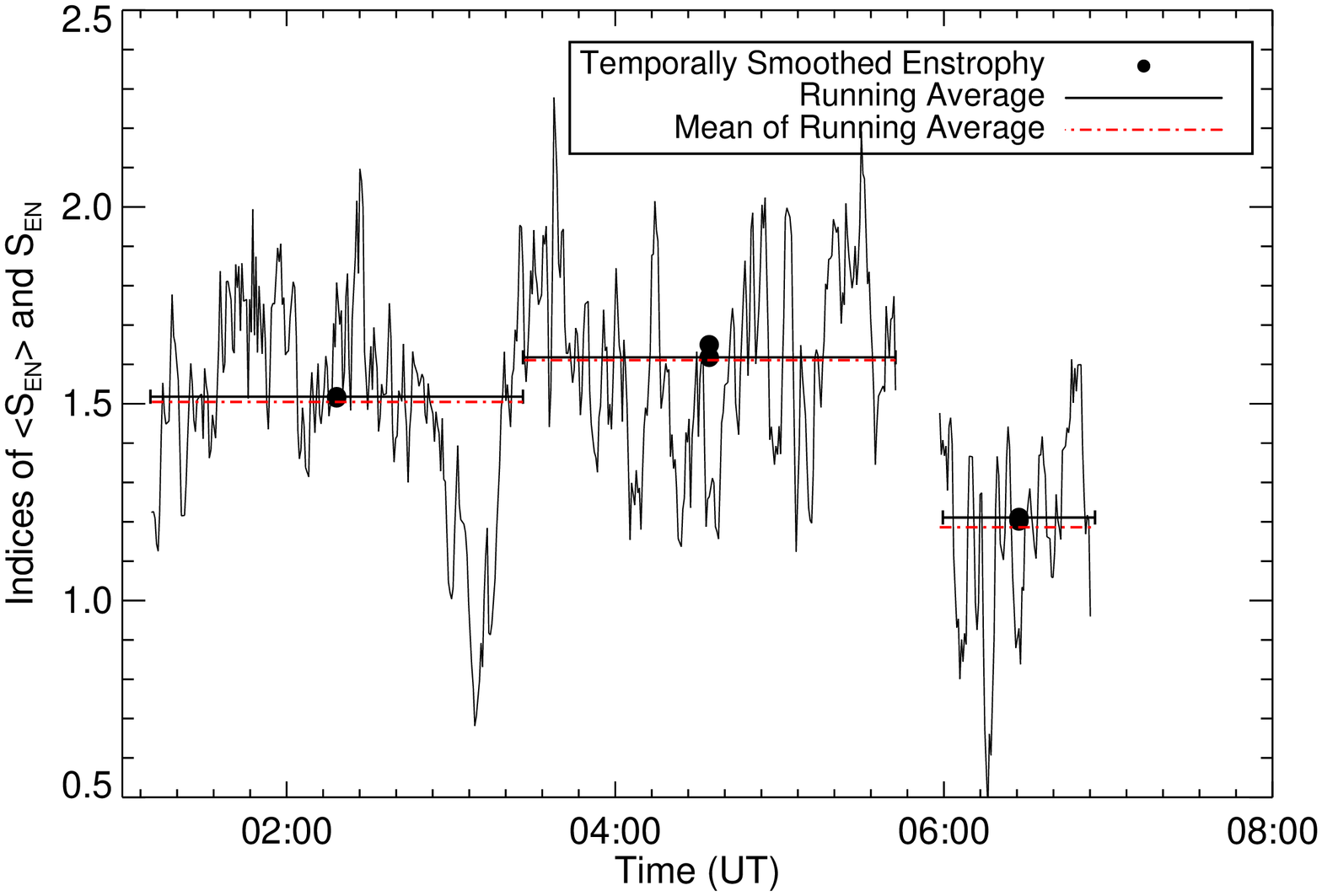}\\
   \includegraphics[angle=90,scale=0.34]{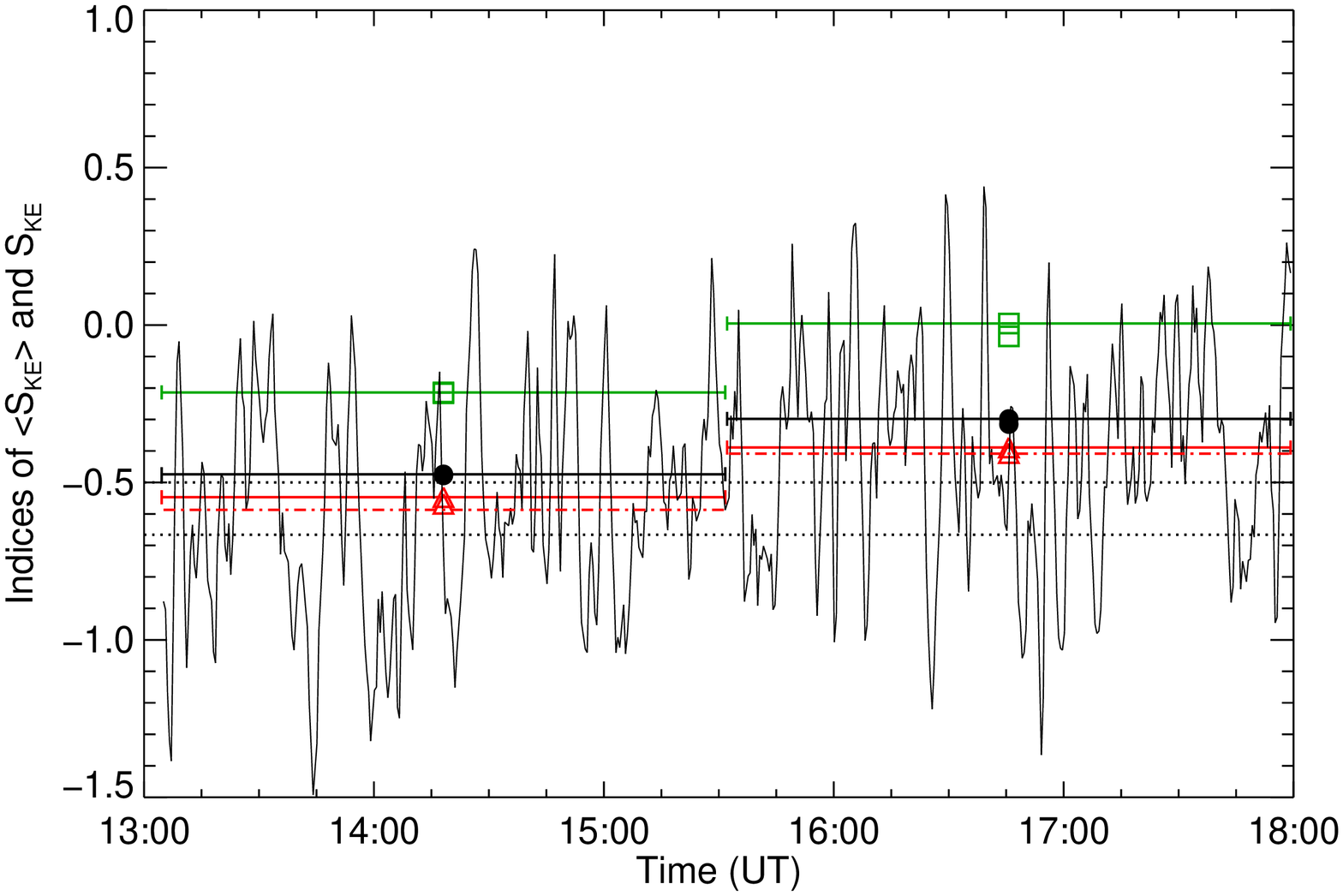} & \includegraphics[angle=90,scale=0.34]{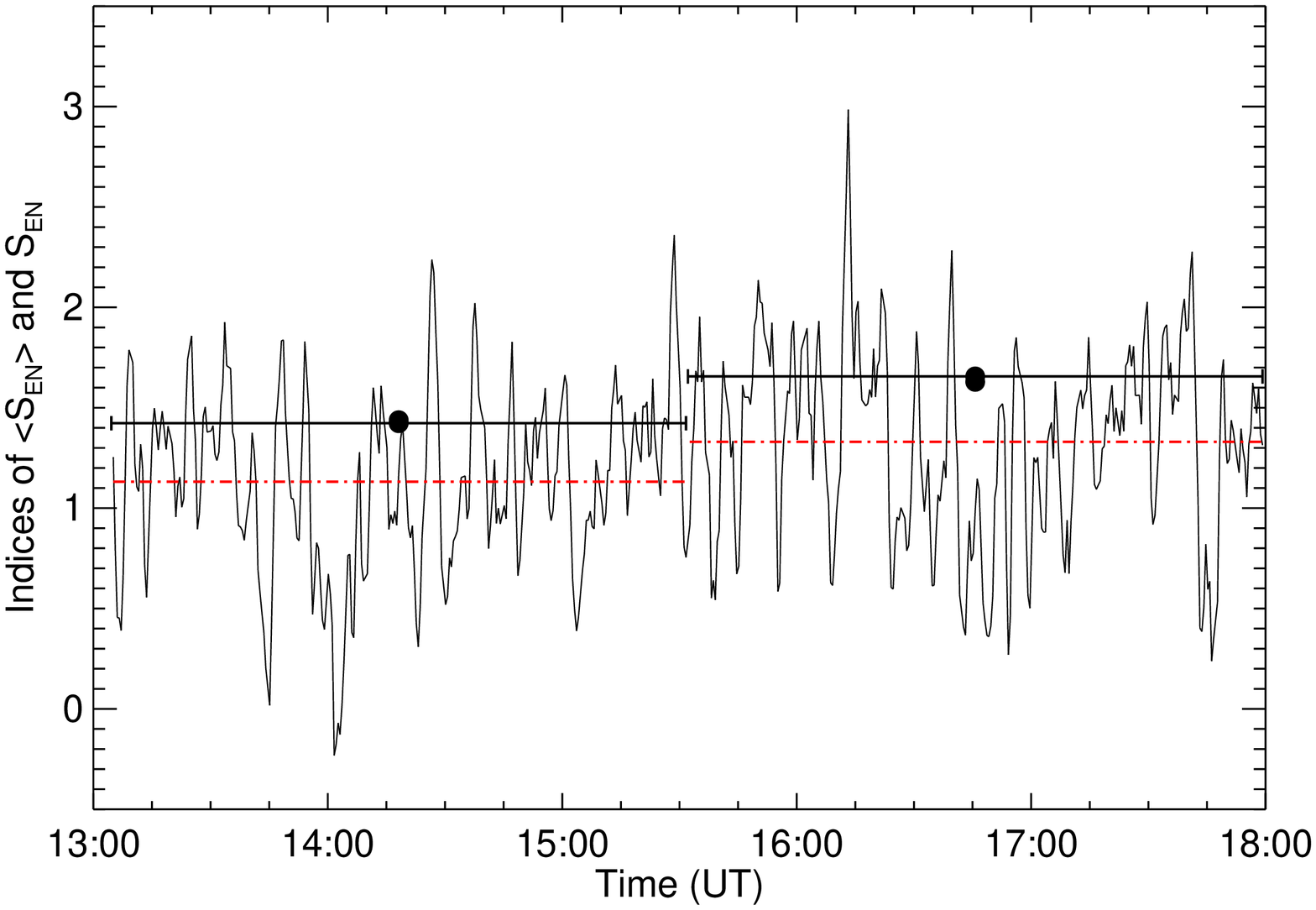}\\
   \includegraphics[angle=90,scale=0.34]{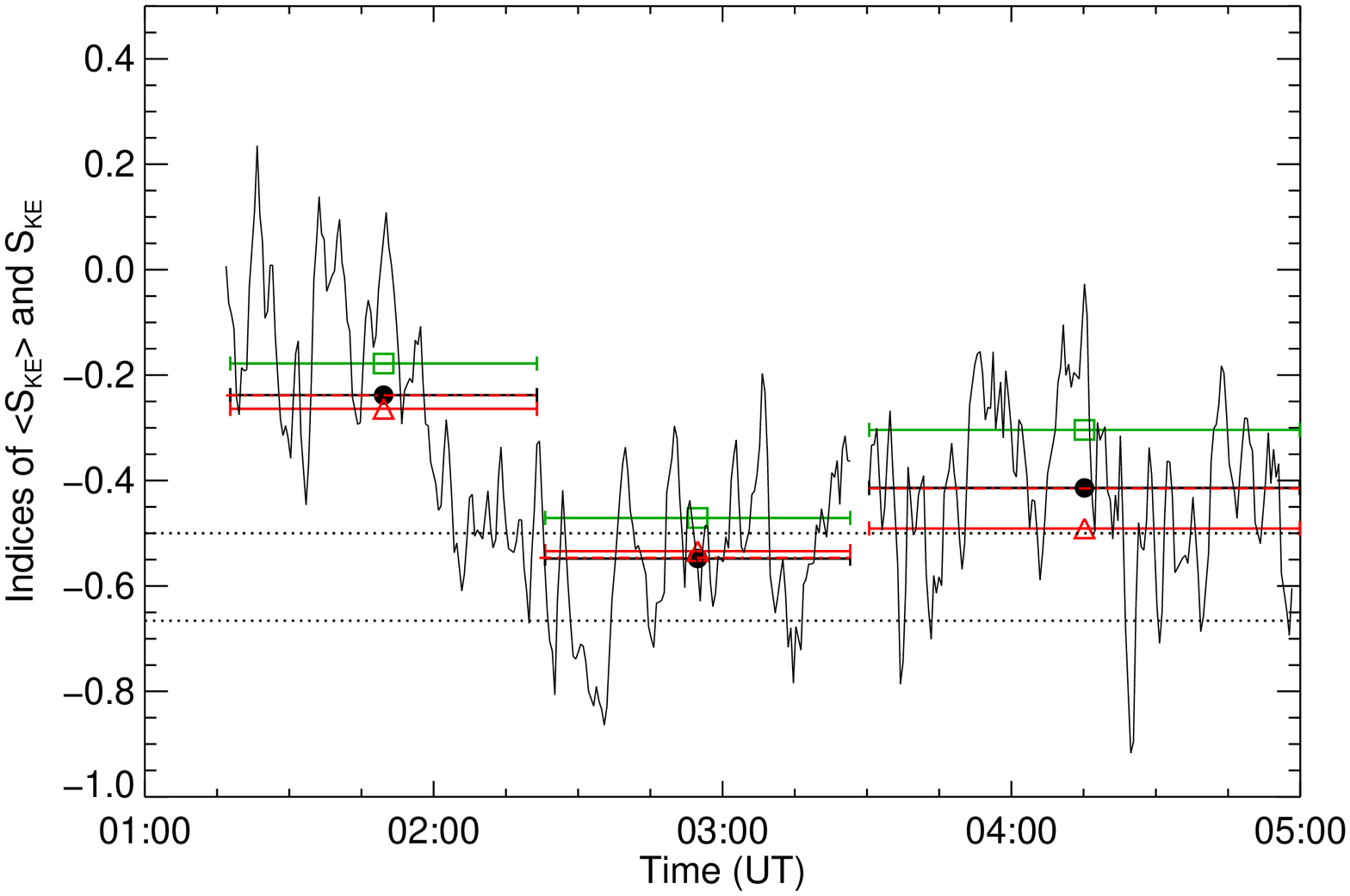} & \includegraphics[angle=90,scale=0.34]{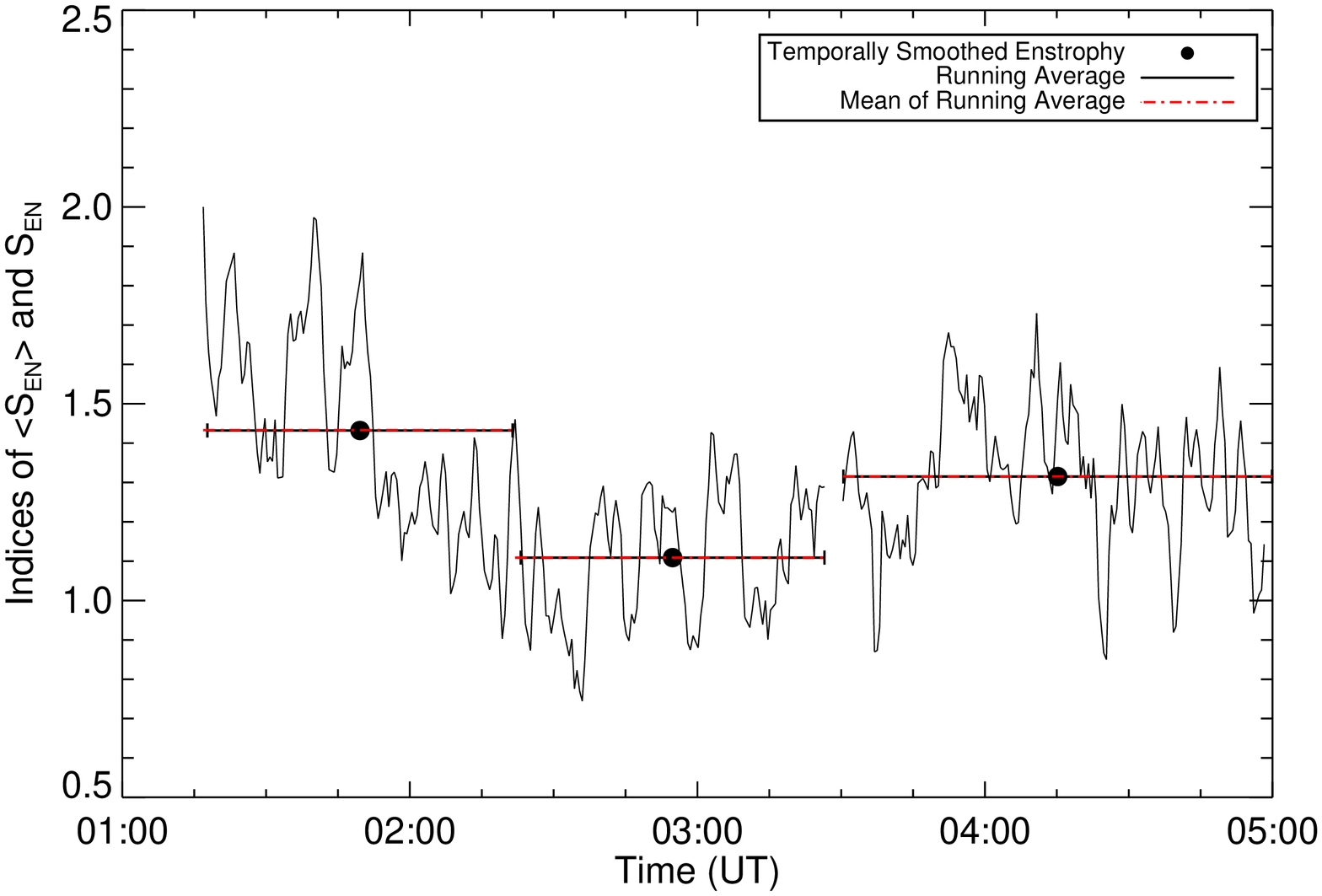}\\
  \end{tabular}
  \caption{The temporal evolution of the indices for $S_{KE}(k_{j})$ (left column) and $S_{EN}(k_{j})$(right column) for the even-numbered prominence frames on 2006/11/30 (top row), 2007/04/25 (middle row), and all frames on 2007/10/03 (bottom row). These plots also show $\left \langle S_{KE}(k_{j}) \right \rangle$, which was calculated from the total/both ($\mathbf{v(x)}$) velocities, horizontal ($u(\mathbf{x})$), and vertical ($v(\mathbf{x})$) velocity components. The horizontal bars indicate the time span used of calculating the temporal averages. There are two data measurements for $\left \langle S_{KE}(k_{j}) \right \rangle$ and $\left \langle S_{EN}(k_{j}) \right \rangle$ over a given time ranges on the 2006/11/30 and 2007/04/25, because these were split into even and odd frames. The horizontal dotted lines in the plots on the left column indicate indices related to different turbulence models, as described in the text.}
  \label{psd_wrt_figure}
\end{figure}

\begin{figure}
\centering
  \begin{tabular}{@{}c@{}}
   \includegraphics[angle=90,scale=0.5]{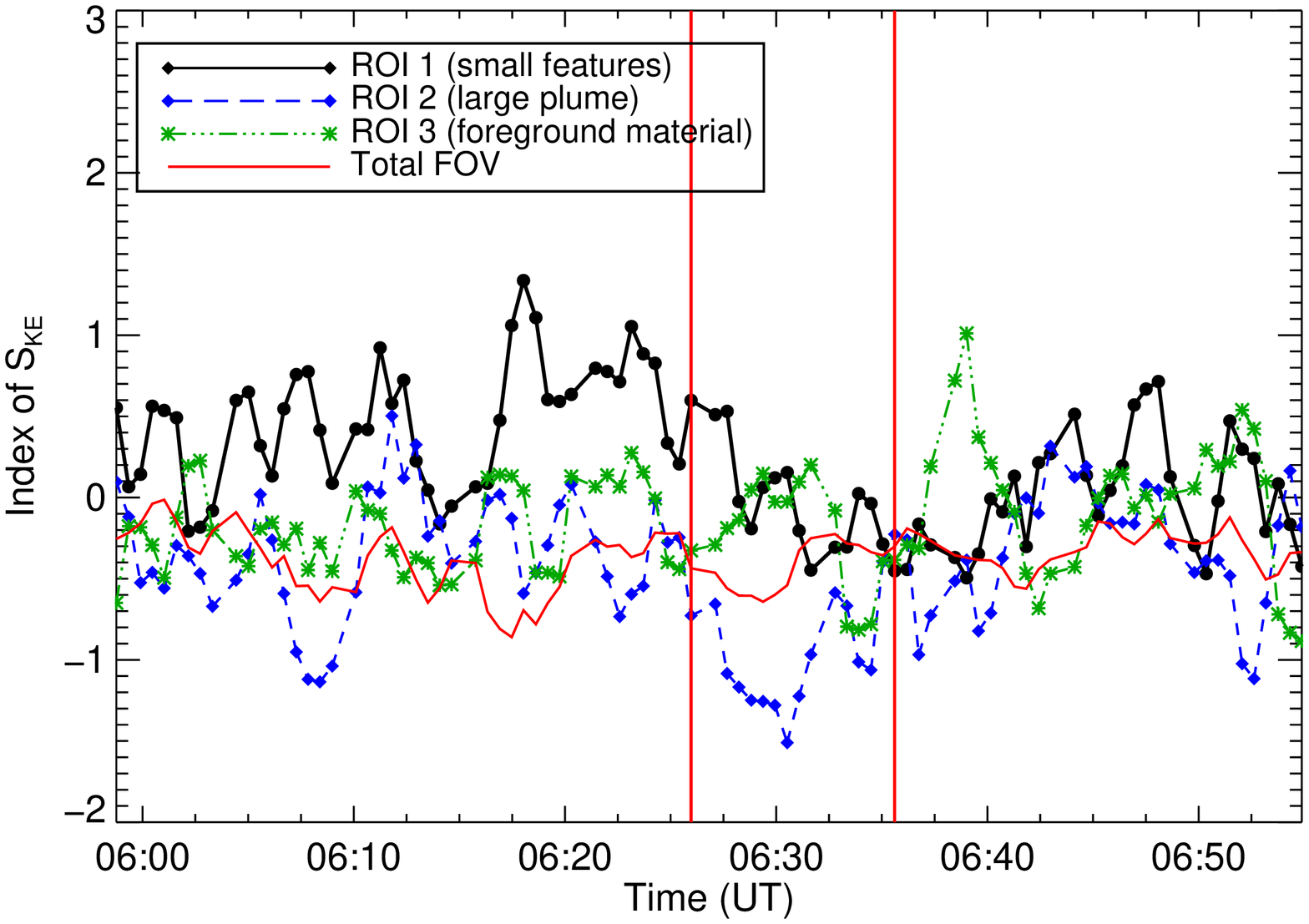}\\
   \includegraphics[angle=90,scale=0.5]{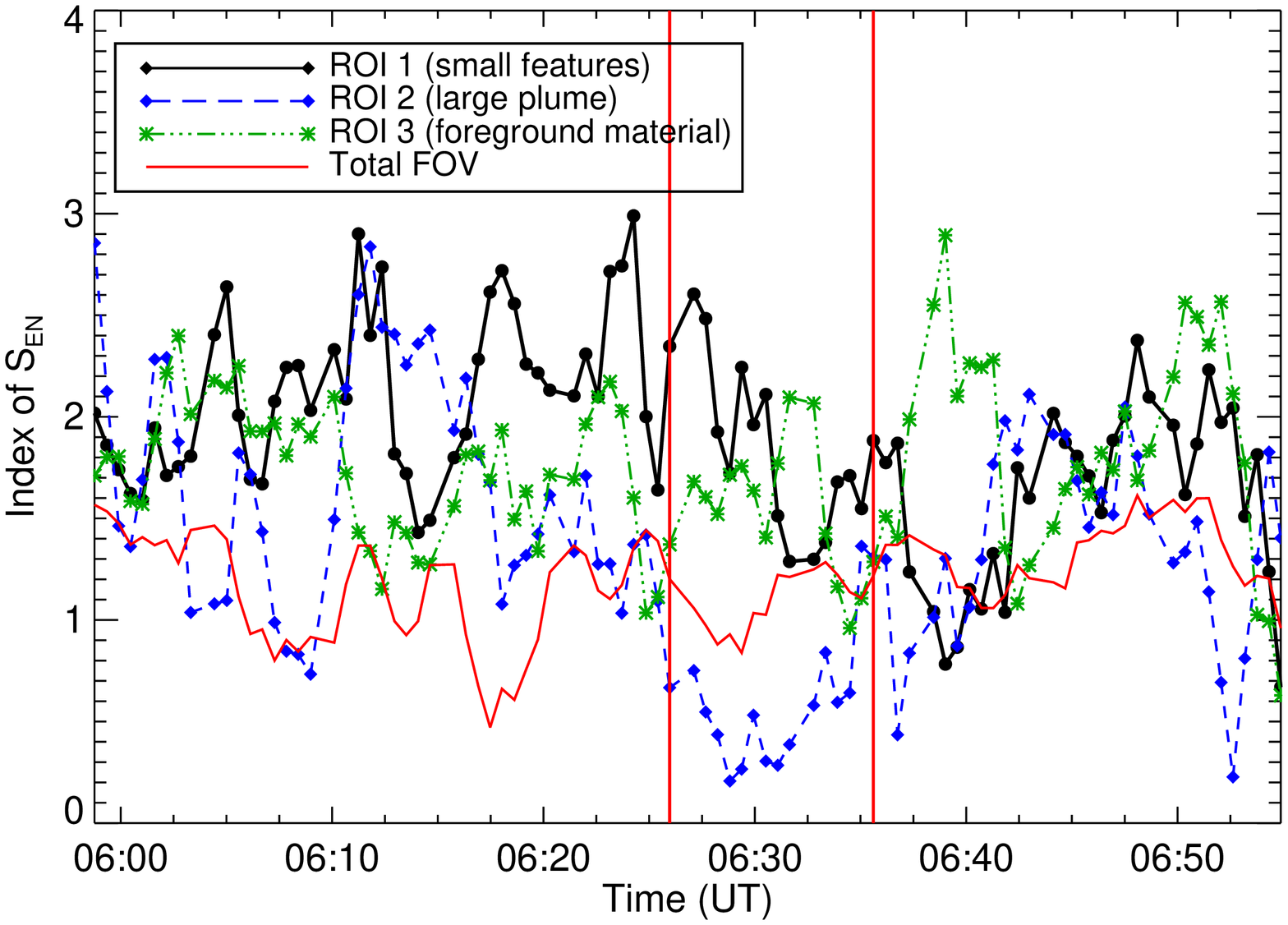}\\
  \end{tabular}
  \caption{ $S_{KE}(k_{j})$ (left column) and  $S_{EN}(k_{j})$ (right column) indices plotted with respect to time for the different Regions Of Interest (ROIs) associated with the last even data set of the 2006/11/30 prominence (shown in Figure \ref{intensity_figure}). The indices found for the entire field of view are shown for comparison by the red line. Additional, red vertical lines are included to indicate when a considerably large plume is crossing into the second ROI's field of view.}
  \label{ROI_KE_enstrophy}
\end{figure}

\begin{figure}
\centering
   \includegraphics[angle=90,scale=0.60]{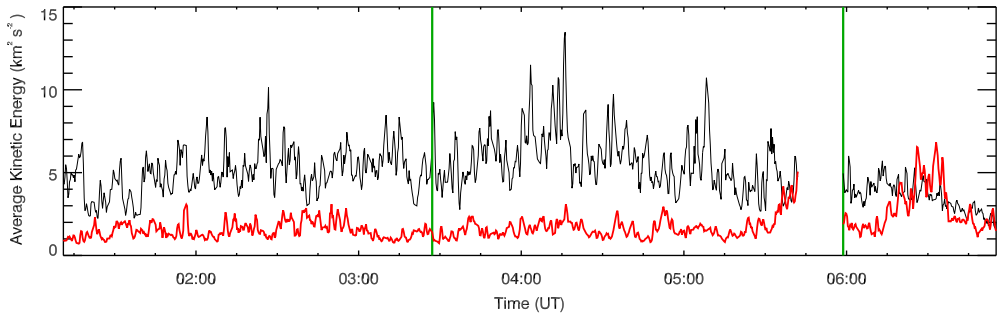}\\
  \caption{This figure shows how the average kinetic energy density ($\epsilon$) changes over time for the 2006/11/30 prominence ROI 1 (black line) and ROI 2 (red line). The even and odd frames were combined for performing this calculation. There is a noticeable increase in $\epsilon$ for ROI 2 when a large plume moves through the FOV around 06:30 UT. $\epsilon$ was calculated using the wavenumber range illustrated in Figure \ref{figure_KE_and_enstrophy_PSD} for performing the spectrum fits and therefore represent a lower limit. Both ROIs illustrated here are $128\times128$ pixels in size. The vertical green lines indicate where data set was divided.}
  \label{ROI_energy_plot}
\end{figure}

\begin{figure}
\centering
   \includegraphics[scale=0.7]{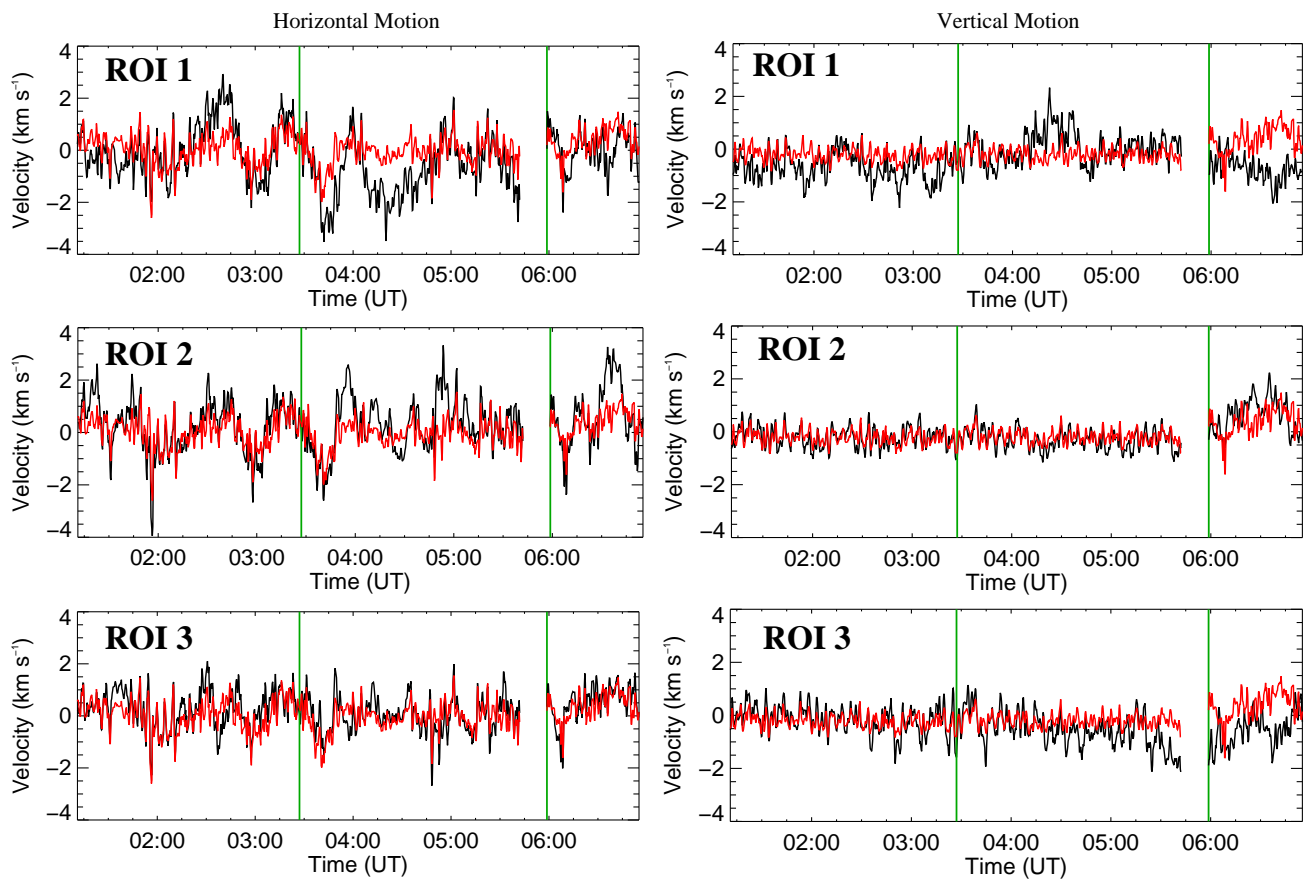}\\
  \caption{This shows how the centroid ($\bar{x}$) of the Gaussian fits used in Figure \ref{histogram_of_velocities} can vary with respect to time. The data only shows the results from the horizontal and vertical velocities associated with the 2006/11/30 prominences, because it was the only prominence that showed any quasi-periodic oscillation. The red line in each plot illustrates how the entire field of view changes in comparison to the smaller Regions Of Interest (ROIs) shown in the top image of Figure \ref{intensity_figure}. The even and odd frames were combined for performing this calculation. The vertical green lines indicate where data set was divided.}
  \label{x_velocity_center}
\end{figure}

The velocity histograms shown in Figure \ref{histogram_of_velocities} were examined further to determine how non-Maxwellian the distributions were. The blue line overlaid shows a Gaussian fit found by a non-linear least-squares fit with three free parameters: $\bar{x}$, $\tilde{\sigma}$, and the amplitude. The mean ($\bar{x}$) or center value and standard deviation ($\tilde{\sigma}$) of the distribution are indicated in the top left corner of Figure \ref{histogram_of_velocities}. The temporally averaged $\tilde{\sigma}$ for each prominence and ROI are listed in Table \ref{KE_enstrophy_values}. The variation of the mean with respect to time was examined for the horizontal and vertical velocity distributions for all of the QPs. However, only the changes to the velocity distribution of the prominence on 2006/11/30 are presented here, because it was the only one to show some interesting quasi-periodic oscillations over time. The red line in Figure \ref{x_velocity_center} shows the oscillation of the horizontal and vertical velocity centroids for the entire FOV from the Gaussian fit. These results are then compared to the horizontal and vertical velocity oscillations seen in our three ROIs. ROI 3 closely matches the results found from examining the entire FOV; the greatest departure is seen when comparing ROI 1 to the entire FOV. There are significant shifts in the horizontal  and vertical velocity distribution in ROI 2 when the large plume enters the FOV. All of the plots in Figure \ref{x_velocity_center} were boxcar smoothed with a width of 3 time steps. We emphasize that extensive care was taken to remove all possible spacecraft jitter in the intensity images before calculating velocities with the FLCT program. Also, the smaller sub-regions display different oscillations when compared to the entire FOV. Any oscillation due to temperature variation of the instrument or error in aligning the data should result in the entire prominence showing the same periodic motion. Thus we believe that the oscillation is not an artifact but an actual feature in the velocities.\par

Oscillations have been reported in prominences to occur at large velocity amplitudes (v $>20$ km s$^{-1}$) due to solar flare activity \citep{Ramsey66}. However, smaller velocity amplitudes (v $\approx2$ km s$^{-1}$) are also frequently seen in filament threads \citep{Lin09}. The periodic motion in the horizontal direction has been linked to granular motions \cite[See Figure 5c from][]{Hillier13a}. Our results could potentially be used to conduct an additional study on the magneto-seismology of the prominence, but we leave this as a discussion of future work.\par

The total kinetic energy ($\epsilon$) and enstrophy per unit mass ($\omega$) were calculated by using Equations (\ref{total_energy_eq}) and (\ref {total_enstrophy_eq}). These values were determined by only including the k-range used for the power-law fits listed in Table \ref{table:C}, because FLCT velocities outside this range are unreliable. The limited range however, results in a loss of information at high wavenumbers and therefore the values reported here are merely a lower limit. The temporal mean and standard deviations of these values are given in Table \ref{KE_enstrophy_values} for all of the prominences and ROIs. The time evolution of $\epsilon$ is also examined for ROI 1 (black line) and ROI 2 (red line) in Figure \ref{ROI_energy_plot} after boxcar smoothing with a window of 3 time steps in width. This shows that there is an increase in $\epsilon$ associated with the large plume entering the FOV, which triggers the release of potential energy stored from having a less dense plasma located below a greater dense plasma. Surprisingly, the largest kinetic energy is associated with what appears to be the quieter ROI 1 on average, which does not show any large disturbances in the intensity movies. However, there are some pixels in ROI 2 where the FLCT did not assigned a velocity to, which could be a contributing factor to this measurement. \par

Equation  (\ref{total_energy_eq}) can also be used to check the spectral density results. First, we calculate $\epsilon$ from the mean-square velocity of a single frame (zero velocities were included in this calculation) and then the powers of the angle-integrated spectra were computed over the entire k-range for comparison. This ratio should equal unity. The temporal mean ratio of $\epsilon$ was then calculated by using the mean-square values over angle-integrated power square were 1.8$\pm$0.4 for 2006/11/30, 1.5$\pm$0.6 for 2007/04/25, and 1.5$\pm$0.3 for 2007/10/03. The consistency of enstrophy can also be checked by using the following 
\begin{equation}
S_{D}(k_{j})+S_{EN}(k_{j})=k_{j}^{2}\; S_{KE}(k_{j})
\label{enstrophy_check}
\end{equation}
where $S_{D}(k_{j})$ comes from the square of the divergence of the velocity field defined as
\begin{equation}
S_{D}(k_{j})=\frac{1}{\Delta K_{j}}\sum_{\mathbf{k} \in A_{j}}\left [ k^{2}_{x}\left |  \tilde{u}_{\mathbf{k}}\right |^2+k^{2}_{y}\left |  \tilde{v}_{\mathbf{k}}\right |^2+2k_{x}k_{y}\mathrm{Re}(\tilde{u}_{\mathbf{k}}\tilde{v}^{*}_{\mathbf{k}})\right ].
\end{equation}
The ratio of the left side of Equation (\ref{enstrophy_check}) to the right were 1.1$\pm$0.1 for 2006/11/30, 1.0$\pm$0.0001 for 2007/04/25, and 1.0$\pm$0.0001 for 2007/10/03 when using the entire k-range in the calculations. It should be noted that the ratio used for checking enstrophy is the result of using only quantities associated with the Fourier transform of the FLCT velocities and they agree more than the ratio from the kinetic energy spectrum, which compares the average FLCT velocities to their Fourier transform.

\section{Conclusion}

In this study, we used SOT data to implement a local correlation tracking method to derive velocities for three separate contrast enhanced prominences. The 2006/11/30 prominence offers an examination of how different regions in a prominence can have varying characteristics, and the 2007/04/25 prominence shows how this methodology works on a less resolved H-$\alpha$ data set. The prominence on 2007/10/03 is an excellent example of a system undergoing a large disturbance or driving force and then moving toward an equilibrium state. The derived velocities were then used to make measurements of the PSDs associated with the kinetic energy and enstrophy for the entire field of view. The objective of this work was to quantify these characteristics to better understand how these prominences evolve and resist erupting.\par 

The intensity images for all three prominences show a break in their power law around $k\approx 3-5 \; \textrm{rads Mm}^{-1}$, which is similar to the range found by \cite{Leonardis12}. It is unclear if this break indicates a change in the physics taking place at different length scales or merely a noise component contributing to the spectrum at small wavenumbers. We feel confident that the presence of the spectral break is not an artifact of the contrast enhancement. \par

An examination of the two-dimensional kinetic energy PSDs, for the velocities normal to the line of sight, showed spectra that were harder then those produced from the intensity images. There is a definite difference in the indices found for $\left \langle S_{KE}(k_{j}) \right \rangle$ in the horizontal and vertical directions, which indicates some anisotropy in the velocities. This is similar to results found by \cite{Hillier14}, which examined Doppler measurements in quiescent prominences with structure functions. It is believed that this discrepancy can be attributed to the mostly horizontal magnetic field found in these prominences \citep{Leroy84}. The range of indices found from the kinetic energy power spectral density, $\left \langle P_{KE}(k) \right \rangle$, were between -1.00 and -1.60 when using both the horizontal and vertical velocities in the measurement. It is inconclusive as to whether these results indicate an agreement with either the Iroshnikov-Kraichnan and Kolmogorov turbulence models. This self-similar behavior existed over the inertial range of $0.8 \leq k\leq (2.0-3.2)\; \textrm{rads} \: \textrm{Mm}^{-1}$ for both kinetic energy and enstrophy. We are not able at present to probe smaller length scales, but we consider it very likely that the inertial range extends to higher k.\par

Enstrophy is calculated for the first time for prominences to quantify the eddies seen in many observations. The enstrophy power spectral density, $\left \langle P_{EN}(k) \right \rangle$, produced indices ranging from +0.11 to +0.65. There is noticeable increase in enstrophy as length scales decrease, which could be an indication of vortex stretching although it is difficult to confirm such a speculation with the two-dimensional velocities resulting from local correlation tracking. A lower limit of the total kinetic energy density ($\epsilon$) $\geq$  0.5-7.1 km$^{2}$s$^{-2}$ and total enstrophy density ($\omega$)  $\geq$  $0.5-13.7\times 10^{-6}\; s^{-2}$ were also calculated from their PSDs.\par

A closer study of three different spatial locations in the 2006/11/30 prominence was also conducted to see how various features can affect the PSDs calculated for the entire FOV. One region (ROI 2) had a large plume enter the FOV. The system responded to this by producing a softer kinetic energy spectrum and a considerable increase in $\epsilon$. Another region (ROI 1) appears quiet in the intensity movies, but displays larger speeds in the velocity maps. This in turn caused the region to have the largest kinetic energy density and hardest spectra. A third region (ROI 3) illustrates how the results can vary when foreground material is present. These sub-regions also showed varying degrees of quasi-periodic oscillations in the vertical and horizontal velocity distributions. This highlights the potential for using these velocities for performing additional seismology studies in the future.\par

An extensive study of different optical flow techniques by \cite{Chae08} has shown that the nonlinear affine velocity estimator (NAVE) \citep{Schuck05} performs superior to LCT programs, but it is slower than the FLCT by two orders of magnitude. This study went with the faster FLCT, but future work in production will explore the differences found when using NAVE or Differential Affine Velocity Estimator (DAVE) \citep{Schuck06} method for determining velocities.\par 

Research by \cite{Lin03} combined stack plots and Doppler measurements to determine the three-dimensional motion of filament threads inside prominences. We see promise in continuing this work by examining velocities in the plane of the sky with local correlation tracking and incorporating high resolution Doppler measurements like those by \cite{Hillier14}. In principle it maybe possible to generate velocity vector maps for the entire quiescent prominences and to then use these maps to analyze turbulence in three dimensions. Recent work by \cite{Schmieder14} also shows potential for generating PSDs of the magnetic field associated with these prominences. Understanding how the magnetic field, kinetic energy, and observed eddies evolve in time are critical to our understanding of how these structures lose stability and erupt. \par

%
\section{\changes{Appendix - Uncertainty of FLCT Velocities}}
\label{Appendix}

In order to ascertain the uncertainty in the FLCT velocity results, two different experiments were performed on synthetic datasets. The first experiment investigates how the ROI and the physical size of features in an image might affect the found FLCT velocities. The second test follows up by attempting to correlate how the difference between the FLCT and known velocities might vary as a function of the physical size of features in an image and by the magnitude of the known velocities. The second experiment is then repeated on a synthetic dataset that closely resembles the characteristics of the first-even dataset associated with the prominence on 2006/11/30. The synthetic images and velocity maps were created in the manner described in Section \ref{Kinetic Energy and Enstrophy PSDs} for testing the reliability of the FLCT program for producing kinetic energy and enstrophy PSDs.

\begin{figure}
\centering
   \includegraphics[scale=0.7]{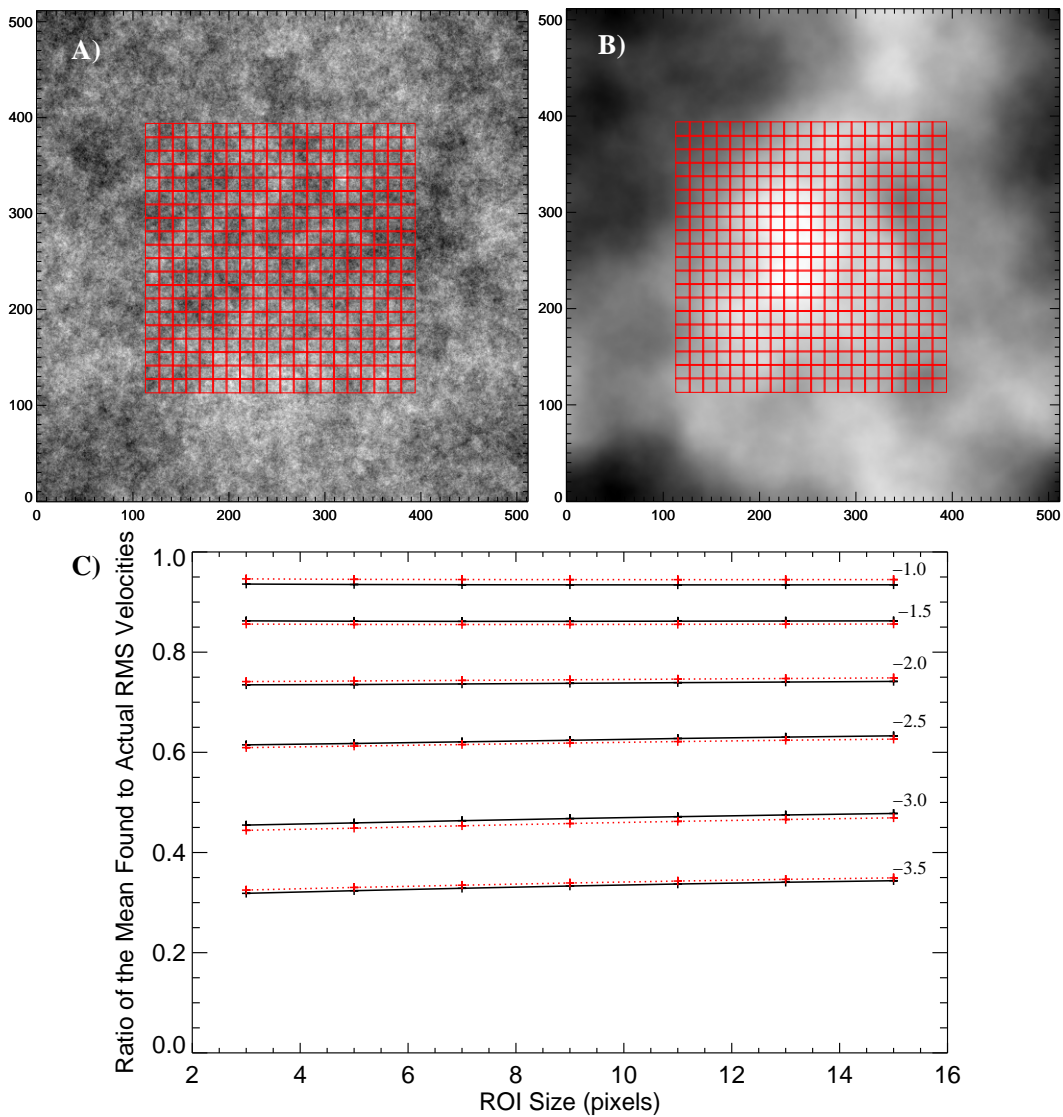}\\
   \caption{Six synthetic images, with varying structure sizes, were created for testing how the ROI might affect the ratio of FLCT velocities to their known values. Figures  \ref{Figure_of_FOV_mean_velo}A and \ref{Figure_of_FOV_mean_velo}B are examples of two intensity images with a power-law index of -1.0 and -3.5 for $S_{I}$, respectively. There is a grid overlaid on top of each image indicating the initial two hundred $15\times15$ pixel ROIs. The different structure sizes were created by varying how soft the intensity spectrum ($S_{I}$) was when creating the intensity images. Figure \ref{Figure_of_FOV_mean_velo}C shows how the ratio of the FLCT to known velocities change as the initial ROI decreases. The solid black lines indicate the horizontal motion and the dashed red lines are for the vertical motion. The power-law indices used for creating each image are stated to the right of each line.}
     \label{Figure_of_FOV_mean_velo}
\end{figure}

Six synthetic images, with varying feature sizes, were created for testing how the ROI might affect the ratio of FLCT velocities to their known velocity values. The different structure sizes were created by varying the power-law index of the intensity spectrum ($S_{I}\sim k^{\alpha}$) when creating the intensity images. Figures  \ref{Figure_of_FOV_mean_velo}A and \ref{Figure_of_FOV_mean_velo}B are examples of two intensity images with a power-law index of -1.0 and -3.5 for $S_{I}$, respectively. There is a grid overlaid on top of each $512\times512$ image indicating the initial two hundred $15\times15$ pixel ROIs. Each dimension of the grid box is then incrementally decreased in size by two pixels until reaching a final size of $3\times3$ pixels. All of the horizontal and vertical velocities were created to maintain an index of -2.5 for $S_{KE}$. The FLCT program was implemented with the parameters: $t$=0.0, $\sigma$=9.0, and $k$=0.4. Figure \ref{Figure_of_FOV_mean_velo}C shows how the ratio of (root mean square FLCT) to (known velocities) changes as the ROI decreases. The solid black lines indicate the horizontal motion and the dashed red lines are for the vertical motion. The power-law indices used for creating each image ($S_{I}\sim k^{\alpha}$) are stated to the right of each line. A hundred different synthetic images -- with the same power-law index -- were used for calculating the ratio of rms velocities.\par

There are a few things worth noting about this test. First, the FLCT velocities always underestimate the known velocities. This is something noted by \cite{Verma13} and \cite{Svanda07} as well. Second, there is very little effect on the accuracy of the FLCT results as the ROI decreases. The biggest ROI effect can be seen when an image used a power-law index of -3.5, but the ratio of means still only dropped by 2.5\%. Lastly, the largest uncertainty occurred when the size of the structures in each image increased, i.e., as the spectrum of the intensity image became softer. This means there is less contrast or dynamic range in each apodization window, resulting in a decrease in the noticeable features that can be correlated with the FLCT program. The next thing to explore is how the magnitude of the velocity might alter the amount of uncertainty in our FLCT velocities.\par

\begin{figure}
\centering
   \includegraphics[scale=0.75]{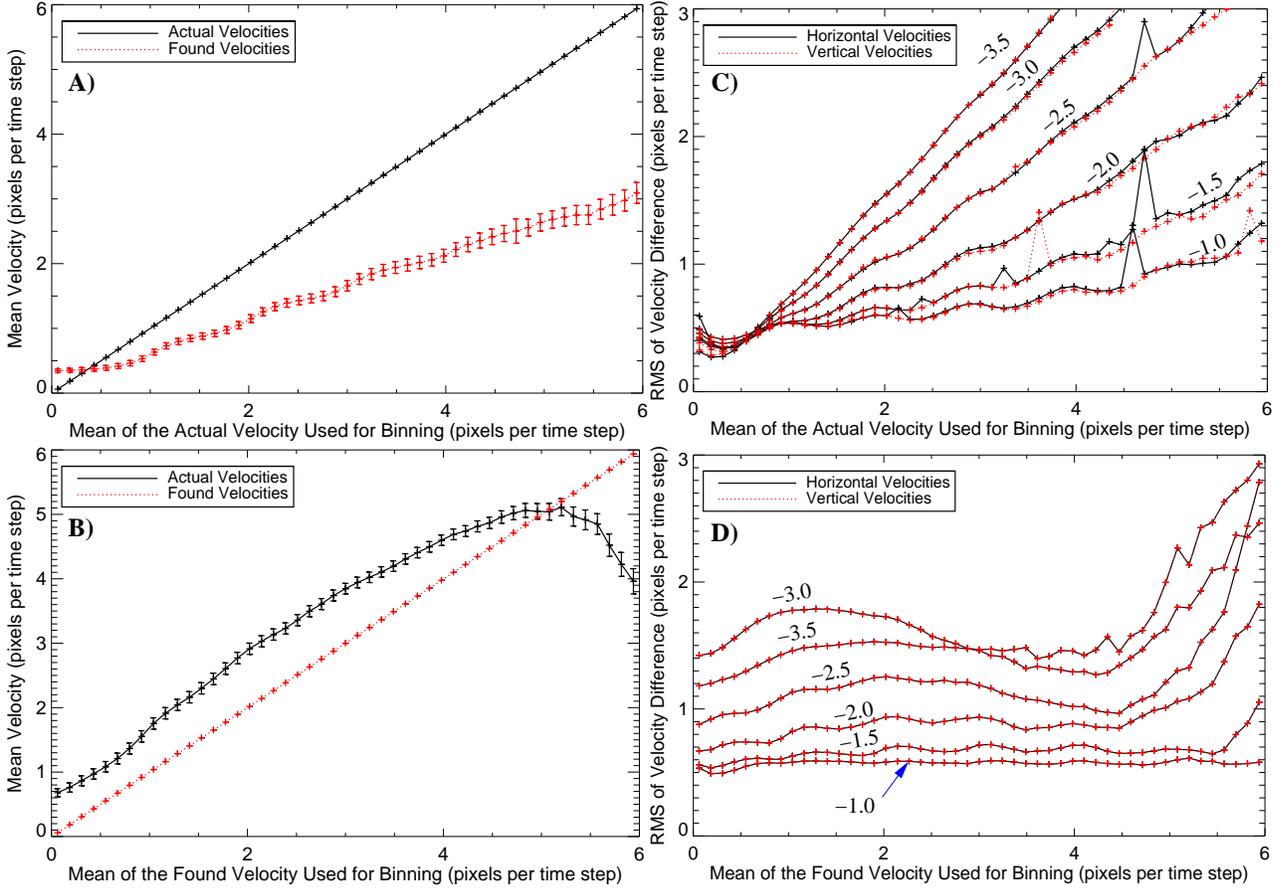}\\
   \caption{These plots show how the uncertainty in the FLCT velocities can vary as a function of the velocity's magnitude and by the size of the structures being tracked. A) illustrates how the found FLCT horizontal velocities are underestimating the magnitude of the actually known horizontal velocities for values above $\approx 0.35$ pixels per time step. B) shows how the FLCT horizontal velocities are lower than the actual velocities for speeds below $\approx 5.0$ pixels per time step. Figure \ref{Fig_act_found_test}A was constructed by binning the actual velocities in the velocity maps and then comparing them to the found velocities at the same spatial locations. The converse of this was done for comparing the FLCT velocities to the known velocities in Figure \ref{Fig_act_found_test}B. The error bars in these figures represent the standard error of the mean. C) and D) goes a step further and shows how the root mean square difference between the actual and found velocities vary as a function of different structure sizes in the intensity image. The power-law indices used for creating each image ($S_{I}\sim k^{\alpha}$) are stated above each line. }
     \label{Fig_act_found_test}
\end{figure}

The greatest advantage of using the synthetic data set is the FLCT velocity maps can be directly compared to the known velocities at each pixel in the image. Our intention now is to determine how well these two values agree as a function of either the known or FLCT velocity magnitudes. This is determined by first identifying the spatial locations of all the known velocity magnitudes in a specified range and then comparing these values to the found velocities at the same location. Figure \ref{Fig_act_found_test}A shows how the found horizontal velocities are once again underestimating the magnitude of the known horizontal velocities for values above $\approx 0.35$ pixels per time step. This is repeated because Figure \ref{Fig_act_found_test}A is only useful if the actual velocities are known a priori. Since in general we do not know the actual velocity, Figure \ref{Fig_act_found_test}B attempts to ascertain how uncertain the FLCT velocities might be when only the derived velocities are known. Figure \ref{Fig_act_found_test}B is also indicating that the FLCT horizontal velocities are lower than the actual velocities for speeds below $\approx 5.0$ pixels per time step. The parameters used for creating the velocity arrays, and for running the FLCT program, are the same as the ones mentioned in the previous experiment. The power law index used for creating the intensity image for Figure \ref{Fig_act_found_test}A and \ref{Fig_act_found_test}B was -2.5. The error bars in these figures represent the standard error of the mean.\par

Figure \ref{Fig_act_found_test}C and \ref{Fig_act_found_test}D goes a step further and shows how the root mean square difference between the actual and found velocities vary as a function of different structure sizes in the intensity image. The power-law indices ($S_{I}\sim k^{\alpha}$) used for creating each intensity image are stated above each line. Figure \ref{Fig_act_found_test}C was constructed by binning the data by the actual known velocities and Figure \ref{Fig_act_found_test}D comes from binning the data by the derived FLCT velocities. All of the data used in Figure \ref{Fig_act_found_test} was constructed by running a hundred different trials with the same input parameters.\par

An examination of Figure \ref{Fig_act_found_test} shows that there is indeed an increase in the error in the derived velocities as the size of the structures in the image increases, i.e., as the power-law indices in $S_{I}\sim k^{\alpha}$ becomes steeper . The plot also indicates that the uncertainty in the FLCT velocity is dependent on the magnitude of the actual input velocity. The difference of the root mean square for the actual velocity and the found FLCT velocities increases approximately monotonically as the magnitude of the actual velocity increases. There appears to be a slight increase in the uncertainty when the measured shifts are below a third of a pixel in size. Ideally a plot like Figure \ref{Fig_act_found_test}C would allow for an estimate of the uncertainty in the FLCT velocities. However, comparing Figure  \ref{Fig_act_found_test}D to \ref{Fig_act_found_test}C illustrates that the uncertainty found when using the FLCT velocities is underestimated for velocities 2-2.5 pixels per time step in size and overestimated for velocities below this value. Despite these noticeable variations in the uncertainty, it may be informative to use this technique on synthetic data that resembles our prominence images.\par

\begin{figure}
\centering
   \includegraphics[scale=0.75]{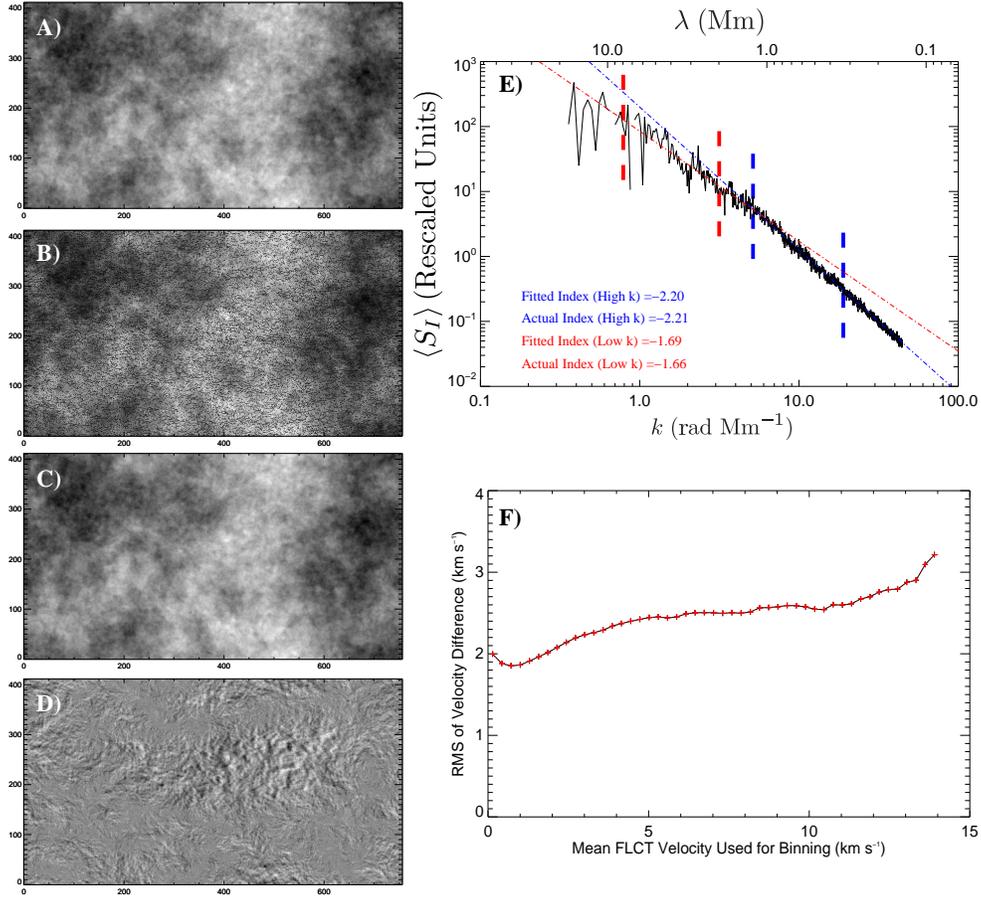}\\
   \caption{ A synthetic dataset was created to resemble the characteristics of the first-even dataset associated with the prominence on 2006/11/30. Figures \ref{Figure_real_data_uncer}A-\ref{Figure_real_data_uncer}D show how the two synthetic images were created for testing the FLCT program. A) is the initial intensity image, which was created from the intensity PSD shown in Figure \ref{Figure_real_data_uncer}E. B) is the image of Figure \ref{Figure_real_data_uncer}A after each pixel has been advected by the known input velocities. This image is then smoothed to produced the second synthetic image shown in Figure \ref{Figure_real_data_uncer}C. The difference between the images in Figure \ref{Figure_real_data_uncer}A \& \ref{Figure_real_data_uncer}C is also given in  Figure \ref{Figure_real_data_uncer}D.  F) illustrates the same quantity given in Figure \ref{Fig_act_found_test}D. This is the root mean square difference between the found and actual velocities when the FLCT velocities are used for the binning. The indistinguishable solid black line indicates the horizontal motion and the dashed red line is for the vertical motion. Figure \ref{Figure_real_data_uncer}F shows that the FLCT velocities have a mean uncertainty corresponding to 2.45$\pm$0.30 km s$^{-1}$ for the prominence on 2006/11/30. }
     \label{Figure_real_data_uncer}
\end{figure}

A synthetic dataset was created to resemble the characteristics of the first-even dataset associated with the prominence on 2006/11/30. Figures \ref{Figure_real_data_uncer}A-\ref{Figure_real_data_uncer}D show how the two synthetic images were created for testing the FLCT program in this appendix and Section \ref{Kinetic Energy and Enstrophy PSDs}. Figure \ref{Figure_real_data_uncer}A is the initial intensity image, which was created from the intensity PSD shown in Figure \ref{Figure_real_data_uncer}E. The indices used for the intensity PSD are the same as the ones shown in Figure \ref{figure_intensity_PSD} with a break in the power-law at $k=0.4$ rad Mm$^{-1}$. The dimensions of the synthetic image are also the same as the prominence observation in units of pixels. Figure \ref{Figure_real_data_uncer}B is the image of Figure \ref{Figure_real_data_uncer}A after the pixels have been advected by the known input velocities. The black spots are locations where the intensity no longer exists due to the advection. These pixels are replaced by the mean value of their surrounding pixels and the resulting second synthetic image after smoothing is shown in Figure \ref{Figure_real_data_uncer}C. The difference between the images in Figure \ref{Figure_real_data_uncer}A \& \ref{Figure_real_data_uncer}C is also given in Figure \ref{Figure_real_data_uncer}D. The actual velocities used for advecting the pixels was constructed to have a power-law index of -1.5 for $S_{KE}$. This comes from the high wavenumber slope shown in Figure \ref{figure_KE_and_enstrophy_PSD}. The low wavenumber slope was not included when constructing the velocities because it resulted in too many pixels being piled up as shown for example in Figure \ref{Figure_real_data_uncer}B. This caused the corresponding smoothed image -- like the one shown in Figure \ref{Figure_real_data_uncer}C -- to be too degraded for performing a proper comparison. The FLCT parameters used were the same as the ones specified for the prominence on 2006/11/30 in Table \ref{table:B}. One hundred such images where created and the difference of the root mean square for the actual velocity and the found FLCT velocities are given in Figure \ref{Figure_real_data_uncer}F. The solid black lines indicate the horizontal motion and the dashed red lines are for the vertical motion. Figure \ref{Figure_real_data_uncer}F shows that the FLCT velocities have a mean uncertainty corresponding to 2.45$\pm$0.30 km s$^{-1}$ for the prominence on 2006/11/30. However, this uncertainty value is higher then the actual uncertainty at low velocity magnitudes as illustrated in Figure \ref{Fig_act_found_test}D when compared to Figure \ref{Fig_act_found_test}C.\par

In summary, it was determined that the FLCT program generally tends to underestimate the velocity magnitudes. The uncertainty in the derived velocities is dependent on the size or more specifically the contrast variation of features in two intensity images being correlated. An increase in the magnitude of the actual velocity can also lead to an increase in the derived velocities' uncertainty. However, since we typically do not know the actual velocity values in our observations, the best estimate we can place on the uncertainty for the prominence on 2006/11/30, i.e., the amount the FLCT velocities underestimate the actual velocity, is a uniform value of approximately 2.45 km s$^{-1}$ from Figure \ref{Figure_real_data_uncer}F. Assuming a constant uncertainty in all velocities would affect the measurements of the total enstrophy and kinetic energy in the research presented here. Therefore, this uncertainty is another reason why the measurements listed in Table \ref{KE_enstrophy_values} and Figure \ref{ROI_energy_plot} are merely a lower limit. Also, the magnitude of the oscillations in Figure \ref{x_velocity_center} and average velocities reported in Table \ref{table:A} are lower due to this uncertainty. We have also presented a method for reporting the uncertainty of the FLCT results for future work, as illustrated in Figure \ref{Fig_act_found_test}.

%
\section{Acknowledgements}
\textit{Hinode} is a Japanese mission developed and launched by ISAS/JAXA, collaborating with NAOJ as a domestic partner, NASA and STFC (UK) as international partners. Scientific operation of the \textit{Hinode} mission is conducted by the \textit{Hinode} science team organized at ISAS/JAXA. This team mainly consists of scientists from institutes in the partner countries. Support for the post-launch operation is provided by JAXA and NAOJ (Japan), STFC (U.K.), NASA, ESA, and NSC (Norway). This work was partially supported by NASA under contract NNM07AB07C with the Smithsonian Astrophysical Observatory, and Living with a Star grant NNX14AD43G. Mikki Wilburn was supported as a participant in the NSF-funded REU program, award number 1156011. The authors would also like to thank Andrew Hillier for his invaluable suggestions and discussions, and the helpful feedback of the anonymous referee. \par
%
{\it Facilities:} \facility{Hinode (SOT)}.

%

 \bibliography{Master_bib_file}  
 
\clearpage

%
%


%

\clearpage

\clearpage

\clearpage

\clearpage

\clearpage

\clearpage

\clearpage

\clearpage

\clearpage

\clearpage

\clearpage

\clearpage

\clearpage


%
%

%
\begin{landscape}
\begin{table}[h]
\centering
\begin{tabular}{ccccccc}
\hline \hline
Solar Object Locator (SOL) & \begin{tabular}[c]{@{}c@{}}Time Span\\ (UT)\end{tabular} & \begin{tabular}[c]{@{}c@{}}Resolution\\ (arcsec/pixel)\end{tabular} & \begin{tabular}[c]{@{}c@{}}Cadence\\ (s)\end{tabular} & Filter Used & \begin{tabular}[c]{@{}c@{}}Horizontal $\left \langle V_{\mathrm{rms}} \right \rangle$\\ (km s$^{-1})$\end{tabular}  & \begin{tabular}[c]{@{}c@{}}Vertical $\left \langle V_{\mathrm{rms}} \right \rangle$\\ (km s$^{-1})$\end{tabular} \\ \hline \hline
SOL:2006-11-30T01:10:47L248C049 & 01:10-06:55 & 0.109 & 17 & Ca\RNum{2} H & 3.81 $\pm$ 0.37 & 4.06 $\pm$ 0.33 \\ \hline
SOL:2007-04-25T13:04:00L120C124 & 13:04-17:59 & 0.16 & 15 & H$\alpha$ & 1.36 $\pm$ 0.25 & 1.55 $\pm$ 0.25\\ \hline
SOL:2007-10-03T01:17:00L156C041 & 01:17-05:00 & 0.109 & 32 & Ca\RNum{2} H & 3.73 $\pm$ 0.41 & 4.47 $\pm$ 0.42 \\ \hline
\end{tabular}
\caption{This is a description of the SOT data used in this study. The temporal average velocities were calculated by removing all zero values, i.e., missing data from the velocity arrays.}
\label{table:A}
\end{table}
\end{landscape}

\clearpage

\begin{table}[h]
\centering
\begin{tabular}{HSSSSSS}
\hline \hline
\multirow{2}{2.0 cm}{\centering Observation Date} & \multicolumn{3}{T}{ Contrast Enhancement Parameters } & \multicolumn{3}{T}{FLCT Parameters} \\ \cline {2-7}
&$\beta$ & Min-Value & Max-Value & $\sigma$ & t & k \\ \hline \hline
2006-11-30 & 50 & 19 & 620 & 9 & 0.075 & 0.4\\ \hline
2007-04-25 & 50 & 3 & 360 & 9 & 0.1 & 0.4\\ \hline
2007-10-03 & 50 & 28 & 700 & 9 & 0.1 & 0.4 \\ \hline
\end{tabular}
\caption{This is a description of the parameters used for contrast enhancing the SOT images and for running the FLCT program.}
\label{table:B}
\end{table}

\clearpage

\begin{landscape}
\begin{table}[h]
\footnotesize
\centering
\begin{tabular}{llllllll}
\hline \hline
\multicolumn{1}{c}{\multirow{2}{*}{\begin{tabular}[c]{@{}c@{}}Nomenclature\\ of Set\end{tabular}}} & \multicolumn{2}{c}{Intensity Index} & \multicolumn{3}{c}{Kinetic Energy Index}  & \multicolumn{1}{c}{\multirow{2}{*}{\begin{tabular}[c]{@{}c@{}}Enstrophy\\ Index\end{tabular}}} & \multicolumn{1}{c}{\multirow{2}{*}{\begin{tabular}[c]{@{}c@{}}Fit Range\\ ($\textrm{rads Mm}^{-1}$)\end{tabular}}} \\
\multicolumn{1}{c}{}    & \multicolumn{1}{c}{\begin{tabular}[c]{@{}c@{}}For Low \\ Wavenumbers\end{tabular}}                                   & \multicolumn{1}{c}{\begin{tabular}[c]{@{}c@{}}\textsuperscript{\textdagger}For High \\Wavenumbers\end{tabular}}               &\multicolumn{1}{c}{\begin{tabular}[c]{@{}c@{}}Horizontal \\ Velocity\end{tabular}} & \multicolumn{1}{c}{\begin{tabular}[c]{@{}c@{}}Vertical\\ Velocity\end{tabular}} & \multicolumn{1}{c}{\begin{tabular}[c]{@{}c@{}}Total\\ Velocity\end{tabular}} & \multicolumn{1}{c}{}   \\ \hline \hline
2006/11/30 Even Set 1 & -1.63(0.07) &   -2.25(0.02)    & -0.21(0.03)   &    0.11(0.03)    &   -0.09(0.01)   &  1.50(0.03) & 0.8-3.2  \\ \hline
2006/11/30 Odd Set 1  & -1.63(0.07) &  -2.25(0.02)     & -0.23(0.03)   &   0.08(0.03)     & -0.12(0.01)     & 1.49(0.03) & 0.8-3.2  \\ \hline
2006/11/30 Even Set 2 & -1.53(0.06) &  -2.30(0.02)     & -0.17(0.03)    &   0.21(0.03)     & -0.02(0.01)     & 1.61(0.03) & 0.8-3.2   \\ \hline 
2006/11/30 Odd Set 2  & -1.54(0.06) &   -2.29(0.02)     & -0.15(0.03)    &   0.21(0.03)     & 0.00(0.01)     & 1.65(0.03) & 0.8-3.2     \\ \hline
2006/11/30 Even Set 3 & -1.24(0.06) &  -2.46(0.01)     & -0.52(0.04)   &  -0.16(0.03)      &  -0.37(0.02)    & 1.20(0.04) & 0.8-3.2   \\ \hline 
2006/11/30 Odd Set 3  & -1.24(0.06) &  -2.45(0.01)     & -0.54(0.03)    & -0.21(0.04)      & -0.41(0.02)     & 1.16(0.04) & 0.8-3.2      \\ \hline
2007/04/25 Even Set 1 & -1.57(0.13) &  -4.06(0.03)    & -0.76(0.10)    & -0.41(0.09)      & -0.59(0.05)     & 1.13(0.14) & 0.8-2.0     \\ \hline   
2007/04/25 Odd Set 1 & -1.57(0.13) &  -4.06(0.03)      & -0.77(0.10)    & -0.42(0.09)      & -0.60(0.05)     & 1.14(0.14) & 0.8-2.0     \\ \hline 
2007/04/25 Even Set 2 & -2.59(0.14) & -3.76(0.03)     & -0.62(0.11)    & 0.18(0.09)     & -0.41(0.06)     & 1.33(0.13) & 0.8-2.0       \\ \hline 
2007/04/25 Odd Set 2 & -2.59(0.14) &  -3.75(0.03)       & -0.64(0.11)    &  -0.22(0.08)    & -0.43(0.06)     & 1.30(0.13) & 0.8-2.0     \\ \hline
2007/10/03 Set 1         & -1.05(0.11) &  -2.61(0.02)      &  -0.26(0.03)   & -0.18(0.03)     & -0.24(0.02)     & 1.43(0.05) & 0.8-3.2     \\ \hline
2007/10/03 Set 2         & -0.86(0.12) &  -2.53(0.03)      & -0.53(0.03)    &  -0.47(0.04)    & -0.55(0.03)    & 1.11(0.07) & 0.8-3.2     \\ \hline 
2007/10/03 Set 3         & -0.94(0.13) &  -2.55(0.03)      & -0.49(0.02)    & -0.30(0.04)     & -0.41(0.03)     & 1.32(0.07) & 0.8-3.2     \\ \hline                                                       
\end{tabular}
\caption{A list that summarizes all of the $\left \langle S(k_{j}) \right \rangle$ indices found. The values in the parenthesis are the uncertainties associated with the fit ($\sigma_{fit}$), given by Equation (\ref{uncertain_eq}). \textsuperscript{\textdagger} The fit range used for producing the intensity indices at high wavenumbers were 5-19 $\textrm{rads Mm}^{-1}$ for 2006/11/30, 3-11 $\textrm{rads Mm}^{-1}$ for 2007/04/25, and 4-12 $\textrm{rads Mm}^{-1}$ for 2007/10/03. All of the other indices where determined from the fit range given in the far right column.  }
\label{table:C}
\end{table}
\end{landscape} 

\begin{table}[h]
\centering
\begin{tabular}{ccccc}
\hline \hline
\multirow{2}{*}{\begin{tabular}[c]{@{}c@{}}Nomenclature\\ of Set\end{tabular}} & \multicolumn{3}{c}{Kinetic Energy Index}    & \multicolumn{1}{c}{\multirow{2}{*}{\begin{tabular}[c]{@{}c@{}}Enstrophy\\ Index\end{tabular}}} \\ 
& \multicolumn{1}{c}{\begin{tabular}[c]{@{}c@{}}Horizontal \\ Velocity\end{tabular}} & \multicolumn{1}{c}{\begin{tabular}[c]{@{}c@{}}Vertical\\ Velocity\end{tabular}} & \multicolumn{1}{c}{\begin{tabular}[c]{@{}c@{}}Total\\ Velocity\end{tabular}} & \multicolumn{1}{c}{}                                                                           \\ \hline \hline
Total Field of View  &-0.52(0.04)    & -0.16(0.03)   & -0.37(0.02)   & 1.20(0.04) \\ \hline
ROI 1   & 0.13(0.06)   & 0.43(0.09)  & 0.23(0.065)   & 1.90(0.12) \\ \hline
ROI 2   & -0.87(0.08)    & -0.01(0.07)  & -0.43(0.06)  & 1.38(0.08)  \\ \hline
ROI 3   & -0.13(0.09)    & -0.03(0.10) & -0.15(0.07)  & 1.75(0.09)   \\  \hline                                                                                                  
\end{tabular}
\caption{These are the indices $\left \langle S_{KE}(k_{j}) \right \rangle$ and $\left \langle S_{EN}(k_{j}) \right \rangle$, associated with different Regions Of Interest (ROI) for the last even data set of 2006/11/30 prominence (shown in Figure \ref{intensity_figure}).}
\label{ROI_results}
\end{table}

\begin{table}[h]
\centering
\label{my-label}
\begin{tabular}{ccccc}
\hline \hline
\begin{tabular}[c]{@{}c@{}}Nomenclature\\ of Set\end{tabular} &\begin{tabular}[c]{@{}c@{}}$\tilde{\sigma_{x}}$ of\\ Gaussian Fit \\ (km s$^{-1}$)\end{tabular} & \begin{tabular}[c]{@{}c@{}}$\tilde{\sigma_{y}}$ of\\ Gaussian Fit\\ (km s$^{-1}$)\end{tabular} & \begin{tabular}[c]{@{}c@{}}Total Kinetic\\ Energy Density\\($\epsilon$) (km$^{2}$ s$^{-2})$\end{tabular} & \begin{tabular}[c]{@{}c@{}}Total Enstrophy\\ Density\\ $(\omega)\times10^{-6}$ (s$^{-2})$\end{tabular} \\ \hline \hline
\begin{tabular}[c]{@{}c@{}}Full FOV\\ 2006/11/30\end{tabular} & 2.19                       & 2.32 & $1.46\pm0.26$      & $2.90\pm0.51$  \\ \hline
\begin{tabular}[c]{@{}c@{}}Full FOV\\ 2007/04/25\end{tabular} & 0.90                       & 0.93 & $0.40\pm0.18$      & $3.8\pm1.90$  \\ \hline
\begin{tabular}[c]{@{}c@{}}Full FOV\\ 2007/10/03\end{tabular} & 2.20                       & 2.71 & $2.37\pm0.5$        & $4.75\pm0.82$  \\ \hline
ROI 1                                                    				       & 2.53                       & 3.24 & $5.05\pm1.99$      & $9.61\pm4.08$  \\ \hline
ROI 2                                                        				       & 1.75                       & 1.90 & $1.80\pm0.97$      & $3.41\pm1.98$   \\ \hline
ROI 3                                                 				               & 1.89                       & 2.47 & $3.07\pm1.44$      & $6.10\pm2.95$   \\ \hline                                                                         
\end{tabular}	
\caption{This table lists the sigma ($\tilde{\sigma}$) for the Gaussian fits illustrated in Figure \ref{histogram_of_velocities}. The total kinetic energy ($\epsilon$) and enstrophy densities ($\omega$) were determined by integrating $\left \langle S_{KE}(k_{j}) \right \rangle$ and $\left \langle S_{EN}(k_{j}) \right \rangle$, respectively, by the limited fit range listed in Table \ref{table:C}. These values constitute a lower limit, since they do not included the whole spectrum.}
\label{KE_enstrophy_values}
\end{table}



\end{document}